\documentclass[twocolumn]{aastex631}
\let\tablenum\relax

\usepackage{amsmath}
\usepackage[separate-uncertainty=true]{siunitx}
\usepackage[version=4]{mhchem}

\DeclareSIUnit{\parsec}{pc}
\DeclareSIUnit{\Lsun}{L_\odot}
\DeclareSIUnit{\Msun}{M_\odot}

\begin{document}

\title{GOALS-JWST: Mid-Infrared Molecular Gas Excitation Probes the Local Conditions of Nuclear Star Clusters and the AGN in the LIRG VV 114}   


\correspondingauthor{Victorine A. Buiten}
\email{buiten@strw.leidenuniv.nl}

\author[0009-0003-4835-2435]{Victorine A. Buiten}
\affiliation{Leiden Observatory, Leiden University, PO Box 9513, 2300 RA Leiden, The Netherlands}

\author[0000-0001-5434-5942]{Paul P. van der Werf}
\affiliation{Leiden Observatory, Leiden University, PO Box 9513, 2300 RA Leiden, The Netherlands}

\author[0000-0001-8504-8844]{Serena Viti}
\affiliation{Leiden Observatory, Leiden University, PO Box 9513, 2300 RA Leiden, The Netherlands}

\author[0000-0003-3498-2973]{Lee Armus}
\affiliation{IPAC, California Institute of Technology, 1200 E. California Blvd., Pasadena, CA 91125, USA}

\author[0000-0003-4909-2770]{Andrew G. Barr}
\affiliation{SRON Netherlands Institute for Space Research, Niels Bohrweg 4, 2333 CA Leiden}

\author[0000-0003-0057-8892]{Loreto Barcos-Muñoz}
\affiliation{National Radio Astronomy Observatory, 520 Edgemont Road, Charlottesville, VA 22903, USA}
\affiliation{Department of Astronomy, University of Virginia, 530 McCormick Road, Charlottesville, VA 22903, USA}

\author[0000-0003-2638-1334]{Aaron S. Evans}
\affiliation{National Radio Astronomy Observatory, 520 Edgemont Road, Charlottesville, VA 22903, USA}
\affiliation{Department of Astronomy, University of Virginia, 530 McCormick Road, Charlottesville, VA 22903, USA}

\author[0000-0003-4268-0393]{Hanae Inami}
\affiliation{Hiroshima Astrophysical Science Center, Hiroshima University, 1-3-1 Kagamiyama, Higashi-Hiroshima, Hiroshima 739-8526, Japan}

\author[0000-0002-1000-6081]{Sean T. Linden}
\affiliation{Steward Observatory, University of Arizona, 933 N Cherry Avenue, Tucson, AZ 85721, USA}

\author[0000-0003-3474-1125]{George C. Privon}
\affiliation{Department of Astronomy, University of Virginia, 530 McCormick Road, Charlottesville, VA 22903, USA}
\affiliation{National Radio Astronomy Observatory, 520 Edgemont Rd, Charlottesville, VA, 22903, USA}
\affiliation{Department of Astronomy, University of Florida, P.O. Box 112055, Gainesville, FL, 32611, USA}

\author[0000-0002-3139-3041]{Yiqing Song}
\affiliation{European Southern Observatory, Alonso de Córdova, 3107, Vitacura, Santiago, 763-0355, Chile}
\affiliation{Joint ALMA Observatory, Alonso de Córdova, 3107, Vitacura, Santiago, 763-0355, Chile}

\author[0000-0002-5807-5078]{Jeffrey A. Rich}
\affiliation{The Observatories of the Carnegie Institution for Science, 813 Santa Barbara Street, Pasadena, CA 91101, USA}

\author[0000-0002-5828-7660]{Susanne Aalto}
\affiliation{Department of Space, Earth and Environment, Chalmers University of Technology, SE-412 96 Gothenburg, Sweden}

\author[0000-0002-7607-8766]{Philip N. Appleton}
\affiliation{IPAC, California Institute of Technology, 1200 E. California Blvd., Pasadena, CA 91125, USA}

\author[0000-0002-5666-7782]{Torsten B\"oker}
\affiliation{European Space Agency, c/o STScI, 3700 San Martin Drive, Baltimore, MD 21218, USA}

\author[0000-0002-2688-1956]{Vassilis Charmandaris}
\affiliation{Department of Physics, University of Crete, Heraklion, 71003, Greece}
\affiliation{Institute of Astrophysics, Foundation for Research and Technology-Hellas (FORTH), Heraklion, 70013, Greece}
\affiliation{School of Sciences, European University Cyprus, Diogenes street, Engomi, 1516 Nicosia, Cyprus}

\author[0000-0003-0699-6083]{Tanio Diaz-Santos}
\affiliation{Institute of Astrophysics, Foundation for Research and Technology-Hellas (FORTH), Heraklion, 70013, Greece}
\affiliation{School of Sciences, European University Cyprus, Diogenes street, Engomi, 1516 Nicosia, Cyprus}

\author[0000-0003-4073-3236]{Christopher C. Hayward}
\affiliation{Center for Computational Astrophysics, Flatiron Institute,
162 Fifth Avenue, New York, NY 10010, USA}

\author[0000-0001-8490-6632]{Thomas S.-Y. Lai}
\affiliation{IPAC, California Institute of Technology, 1200 E. California Blvd., Pasadena, CA 91125, USA}

\author[0000-0001-7421-2944]{Anne M. Medling}
\affiliation{Department of Physics \& Astronomy and Ritter Astrophysical Research Center, University of Toledo, Toledo, OH 43606,USA}
\affiliation{ARC Centre of Excellence for All Sky Astrophysics in 3 Dimensions (ASTRO 3D); Australia}

\author[0000-0001-5231-2645]{Claudio Ricci}
\affiliation{Instituto de Estudios Astrof\'{\i}sicos, Facultad de Ingenier\'{\i}a y Ciencias, Universidad Diego Portales, Avenida Ejercito Libertador 441, Santiago, Chile}
\affiliation{Kavli Institute for Astronomy and Astrophysics, Peking University, Beijing 100871, China}

\author[0000-0002-1912-0024]{Vivian U}
\affiliation{Department of Physics and Astronomy, 4129 Frederick Reines Hall, University of California, Irvine, CA 92697, USA}




\begin{abstract}
The enormous increase in mid-IR sensitivity and spatial and spectral resolution provided by the JWST spectrographs enables, for the first time, detailed extragalactic studies of molecular vibrational bands. 
This
opens an entirely new window for the study of the molecular interstellar medium in luminous infrared galaxies (LIRGs).
We present 
a detailed analysis of rovibrational bands of gas-phase \ce{CO}, \ce{H2O}, \ce{C2H2} and HCN towards the heavily-obscured eastern nucleus of the LIRG VV\,114, as observed by NIRSpec and MIRI MRS
. Spectra extracted from apertures of \SI{130}{\parsec} in radius show a clear dichotomy between the obscured AGN and two intense starburst regions. {We detect the \SI{2.3}{\micron} CO bandheads, characteristic of cool stellar atmospheres, in the star-forming regions, but not towards the AGN.} Surprisingly, at $\SI{4.7}{\micron}$ we find highly-excited CO (${T_\mathrm{ex} \approx 700-\SI{800}{\kelvin}}$ out to at least rotational level $J = 27$) towards the star-forming regions, but only cooler gas ($T_\mathrm{ex} \approx {\SI{200}{\kelvin}}$) towards the AGN. We conclude that only mid-infrared pumping through the rovibrational lines can account for the equilibrium conditions found for CO and \ce{H2O} in the deeply-embedded starbursts. {Here the CO bands probe regions with an intense local radiation field inside dusty young massive star clusters or near the most massive young stars.} 
The lack of high-excitation molecular gas towards the AGN is attributed to geometric dilution of the intense radiation from the {bright point source}. An overview of the relevant excitation and radiative transfer physics is provided in an appendix.
\end{abstract}


\section{Introduction} \label{sec:intro}

The $L_\mathrm{IR} = \SI{4.5e11}{\Lsun}$ luminous infrared galaxy (LIRG) VV\,114 (IC\,1623, Arp\,236) is a merging system undergoing intense star-forming activity. It consists of two gas-rich galaxies, with a projected nuclear separation of $\SI{8}{\kilo\parsec}$ at a distance of \SI{80}{\mega\parsec}. At optical wavelengths, its western component (VV\,114W) shows bright spiral arms and many luminous young star clusters, while its eastern component (VV\,114E) is heavily obscured by dust lanes. At {mid- and far-IR} wavelengths, however, the eastern component dominates the emission. With a \SI{7}{\micron}-to-UV flux density ratio of $\sim 800$ for VV\,114E and $\sim 10$ for VV\,114W \citep{Charmandaris-LeFloc'h2004}, VV\,114 has the most extreme colour contrast between {galaxies taking part in the merger} observed among local LIRGs.

Early radio and infrared images were already able to resolve the dusty eastern nucleus into a northeastern (NE) and southwestern (SW) component \citep{Condon1991, Knop1994}. The latter has a complex structure, consisting of a bright unresolved point-like source and several secondary emission peaks \citep{Scoville2000, Soifer2001}. The point source is bright at infrared wavelengths, but does not coincide with any of the peaks in the \SI{8.4}{\giga\hertz} radio image.
High-resolution ($\theta_\mathrm{FWHM} \approx 0\farcs2$) VLA \SI{33}{\giga\hertz} imaging by \citet{Song2022} resolved the SW core into four distinct components. The \SI{33}{\giga\hertz} continuum, which traces the thermal free-free emission from HII regions, is much fainter at the location of the infrared point source than at both the NE core and at an extended knot south of the infrared peak, suggesting relatively weak star-forming activity {at the point source} \citep{Evans2022}.


Atacama Large Millimeter/submillimeter Array (ALMA) observations reveal abundant cold molecular gas in VV\,114. CO is ubiquitous in the system, while the dense gas tracer HCN is found to be concentrated in the nucleus of VV\,114E; \ce{HCO+} is found in the eastern nucleus and in the overlap region between the two galaxies \citep{Saito2015}. An enhanced methanol abundance, indicative of \SI{}{\kilo\parsec}-scale shocks, is found in the overlap region as well \citep{Saito2017}. High-resolution {0\farcs2 ALMA Band 7 observations of \ce{HCN}(4-3) and \ce{HCO+}(4-3) were able to resolve the VV\,114E nucleus into the NE core and a complex SW structure \citep{Saito2018}, with comparatively little emission coming from the bright infrared point source.}

The nature of the nuclear components of VV\,114E has been a subject of debate. Mid-infrared ISO studies \citep{LeFloc'h2002} and X-ray observations \citep{Grimes2006} found no evidence for an energetically important active galactic nucleus (AGN), but could not exclude {a heavily-obscured one} either. \citet{Iono2013} found a significantly enhanced HCN(4-3)/HCO$^+$(4-3) ratio towards the NE core, and attributed this to the presence of a deeply-embedded AGN. Near- and mid-infrared JWST imaging and spectroscopic observations, however, show a different picture. \citet{Evans2022} found mid-IR colours that are consistent with a starburst towards the NE core, while those towards the SW core are in line with an AGN. Moreover, the extremely low equivalent widths of \SI{3.3}{\micron} and \SI{6.2}{\micron} polycyclic aromatic hydrocarbon (PAH) features found by \citet{Rich2023} are strong evidence {for the presence of an AGN in the SW core.} 
These JWST observations indicate that the IR-bright point source in the SW core harbours an AGN, while the other nuclear continuum peaks are starburst regions.

The integral field spectrographs aboard JWST provide a new method of probing local gas conditions. Their high sensitivity and spatial resolution and moderate spectral resolution enable detailed analysis of rovibrational molecular bands in external galaxies. Towards U/LIRGs, the intense infrared background radiation causes these lines to be seen in absorption, allowing for direct measurements of column density.
{Since the high-$J$ lines will not be affected by cold foreground gas, the excitation conditions of the gas can be studied in detail.} Such studies have previously only been possible in Galactic star-forming regions \citep[e.g.][]{Lahuis-vanDishoeck2000, Gonzalez-Alfonso2002}.

{The Q-branches} of \ce{C2H2}, HCN and \ce{CO2}, {blends of closely-spaced pure vibrational lines}, have been detected by Spitzer towards a handful of local U/LIRGs \citep{Lahuis2007}, but at lower spatial and spectral resolution. \citet{Rich2023} previously reported the detection of the $\nu_2$ \SI{14.0}{\micron} band of HCN and the $\nu_5$ \SI{13.7}{\micron} band of \ce{C2H2} in absorption towards the NE core in VV 114E.
The fundamental band of CO at \SI{4.67}{\micron}, {tracing transitions between the ground state and the first vibrationally excited state}, has been observed in absorption towards several heavily-obscured U/LIRGs from the ground \citep[e.g.][]{Spoon2003, Geballe2006, Shirahata2013, Baba2022, Ohyama2023}. \citet{Pereira-Santaella2023} demonstrated JWST's capabilities for studying the CO band in emission.

In this paper, we present JWST NIRSpec and MIRI absorption spectra of gas-phase CO, \ce{H2O}, \ce{C2H2} and HCN observed towards the NE core and the SW complex in the nucleus of VV 114E. 
Using rotation diagram analysis (CO \& \ce{H2O}) and spectral fitting (\ce{C2H2} \& HCN), we probe the excitation conditions of the molecular gas, and investigate the responsible excitation mechanism.

Adopting a cosmology of $H_0 = \SI{70}{\kilo\meter\per\second\per\mega\parsec}$, $\Omega_m = 0.28$ and $\Omega_\Lambda = 0.72$, the redshift of VV 114 ($z = 0.0202$) corresponds to an angular scale of $\SI{400}{\parsec} / \SI{1}{\arcsec}$.

\section{Observations \& Data Reduction}
\begin{figure}
    \centering
    \includegraphics[width=\linewidth]{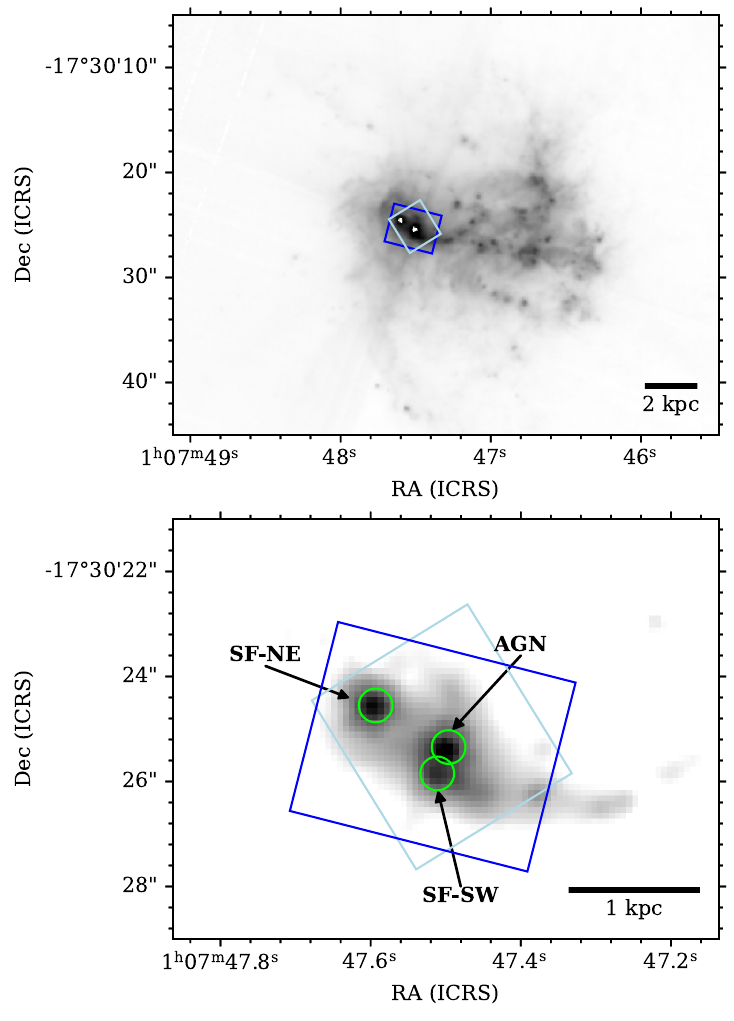}
    \caption{MIRI images of VV\,114, {where North is up and East is to the left}. The field of view of NIRSpec and MIRI Channel 1 are indicated in light blue and dark blue respectively. Top: MIRI F770W image of the full system. In the eastern nucleus, saturated pixels appear as white dots. Bottom: MIRI F770W SUB128 image. The apertures of $0\farcs32$ in radius considered in this work are indicated by green circles. {The association of these regions with an AGN and star-forming regions is motivated by the work of \citet{Evans2022} and \citet{Rich2023} (see text).}}
    \label{fig:miri_image}
\end{figure}

VV\,114E was observed with the JWST Mid-InfraRed Instrument \citep[MIRI,][]{Rieke2015,Labiano2021} in Medium Resolution Spectroscopy (MRS) mode \citep{Wells2015} on July 5, 2022, and with NIRSpec in IFU mode \citep{Jakobsen2022,Boeker2022} on July 19, 2022 as part of the GOALS Early Release Science (ERS) Program 1328 (co-PIs Armus and Evans; \dataset[10.17909/vhw3-g317]{\doi{10.17909/vhw3-g317}}). This work is focused on the spectroscopic data, but we include the results of MIRI imaging to indicate the regions of spectral extraction (Fig.~\ref{fig:miri_image}). For a description of the MIRI imaging data, see \citet{Evans2022}.

\subsection{NIRSpec Data}
NIRSpec IFU observations cover the wavelength range from 0.97 to $\SI{5.27}{\micron}$, employing three combinations of gratings and filters: G140H/F100LP, G235H/F170LP and G395H/F290LP. This last configuration covers the fundamental band of CO at $\SI{4.7}{\micron}$. A four-point dither pattern was used, as well as dedicated \textit{Leakcal} images to correct for failed open MSA shutters. The field of view (FOV) of the data products is $\approx 3\farcs6 \times 3\farcs8$ at a position angle (counterclockwise from North) $\text{PA} \approx \SI{32}{\degree}$.

We downloaded the uncalibrated data---processed by SDP version 2023 1a---from the MAST portal.
We reduced the data using JWST Science Calibration Pipeline version \texttt{1.11.1.dev1+g9975346} and CRDS reference file \texttt{jwst\_1097.pmap}. {This is an updated version of the pipeline with respect to the one applied by \citet{Rich2023}.}

Both the science and Leakcal exposures were first put through the \texttt{Detector1} pipeline, which applies detector-level calibrations to produce count rate files from nondestructive ramp readouts. The \texttt{Spec2} pipeline then processes these rate files to produce fully calibrated individual exposures. At this step, additional instrumental corrections are applied, as well as flux and wavelength calibration. The Leakcal count rate files are used here to correct for failed open MSA shutters. The \texttt{Spec3} pipeline finally combines the Stage 2 calibrated science exposures into a fully calibrated data cube. Stage 3 processing includes an outlier detection step. After some experimentation, we decided on an outlier flag threshold of 99.8\% and a kernel size of $7 \times 7$. The final data cube spaxel size is $0\farcs1$, corresponding to $\approx \SI{40}{\parsec}$ for VV\,114. {The NIRSpec PSF has a FWHM of $\sim 0\farcs17$ at the long-wavelength end of $\SI{5.3}{\micron}$ \citep{Jakobsen2022}.}

\subsection{MIRI MRS Data} \label{sec:miri_reduction}
The MIRI MRS observations cover the full $4.9 - \SI{27.9}{\micron}$ wavelength range of the four IFU channels, using the three grating settings SHORT (A), MEDIUM (B) and LONG (C) to do so. Dedicated off-target background observations were taken, and a four-point dither pattern was employed. Each channel has a different FOV; it is smallest for channel 1 with a FOV of $\sim \SI{5}{\arcsec} \times \SI{4}{\arcsec}$ at a position angle $\text{PA} = \SI{255}{\degree}$. {The FWHM of the MRS PSF varies between 0\farcs35 in channel 1A and 0\farcs9 in channel 4C \citep{Argyriou2023}.}

The uncalibrated data products---processed by JWST Science Data Processing (SDP) version 2022 4a---were downloaded through the MAST Portal. These were then reduced using developmental version \texttt{1.11.1.dev1+g9975346} of the JWST Science Calibration Pipeline \citep{bushouse_howard_2022_7229890} and calibration reference data system (CRDS) reference file \texttt{jwst\_1094.pmap}.


The MIRI MRS data were processed in a similar fashion to the NIRSpec data. The \texttt{Detector1} pipeline produced count rate files, and we applied the \texttt{Spec2} step to subsequently turn these into calibrated individual exposures. We also applied the residual fringe correction (rfc) at this stage. For the background subtraction, we applied the master background subtraction step in the \texttt{Spec3} pipeline. Outlier detection was performed at this stage as well, with a threshold of 99.8\% and a kernel size $11 \times 1$. Finally, the \texttt{Spec3} pipeline combined the Stage 2 exposures into data cubes.

\subsection{Astrometric corrections}

To ensure that the regions from which we extract spectra are the same between the NIRSpec and MIRI cubes, we calibrate the astrometry to align with Gaia stars, employing the simultaneous MIRI imaging data. We find an offset of 0\farcs2 between the native MIRI astrometry and the detected Gaia stars in the parallel F560W image. After applying this correction to the MIRI MRS cubes, we collapse both the MRS and the NIRSpec cubes in the (observer-frame) 5.0-\SI{5.1}{\micron} spectral region, and shift the NIRSpec astrometry such that the two collapsed images match. The correction required for the NIRSpec cube is found to be 0\farcs3.

\subsection{Spectral extraction}

The nuclear region of VV\,114E contains a large number of mid-IR luminous cores \citep{Evans2022}. One of these is most likely an obscured AGN, as suggested by \citet{Rich2023}. Two other prominent mid-IR peaks display strong, {most likely} thermal radio emission at 33\,GHz \citep{Evans2022, Song2022}, 
and therefore represent regions of intense star formation. Together these three cores dominate the \SI{7.7}{\micron} emission from the VV\,114E nucleus, and they form the subject of study of the present paper. For easy reference, we will henceforth refer to these as AGN, SF-NE and SF-SW, and their positions are indicated in Fig.~\ref{fig:miri_image}. These cores correspond to the ``SW core'', ``NE core'' and ``SW-S knot'' respectively, in the notation of \citet[][their Fig.~3]{Evans2022}. In the notation of \citet{Rich2023}, they correspond closely to apertures c, a and d respectively; they also align with the HCO$^+$ (4-3) emission peaks denoted by E2, E0 and E1 respectively by \citet{Iono2013}.



We extract spectra from apertures of 0\farcs322 (\SI{130}{\parsec}) in radius, centered on the {apparent intensity peaks in the \SI{5.0}{\micron} slice of the NIRSpec data cube}. The position and sizes of the extraction apertures are indicated in Fig.~\ref{fig:miri_image}. {Although the JWST Science Calibration Pipeline contains a 1D residual fringe correction for extracted spectra, in addition to the residual fringe correction performed in the \texttt{Spec2} step (see Section \ref{sec:miri_reduction}),} we do not apply this {1D correction} for the majority of this work, as this defringer can partially remove the rovibrational lines of interest due to their periodic nature. We only apply it for the spectral fits for \ce{C2H2} and HCN (see Section \ref{subsec:analysis-14micron}). {No aperture corrections are applied as 
{our analysis will be based on normalised spectra.}}

To correct for residual systematic errors in the wavelength solution of the instruments, we manually calibrate the redshift of the extracted spectra. We use \ce{H2} emission lines to set the systemic velocity, as these lines trace the bulk of the molecular gas in the nuclear regions. Since the MIRI MRS wavelength solutions are specific to the grating setting used, we only consider H$_2$ lines close in wavelength to the molecular features under consideration, and we do so separately for each species studied. We do not, however, differentiate between the spatial regions, as the difference in radial velocity between regions is 
smaller than the difference in radial velocity between different \ce{H2} lines.




\section{Results}
\begin{figure*}[ht!]
    \centering
    \includegraphics[width=\linewidth]{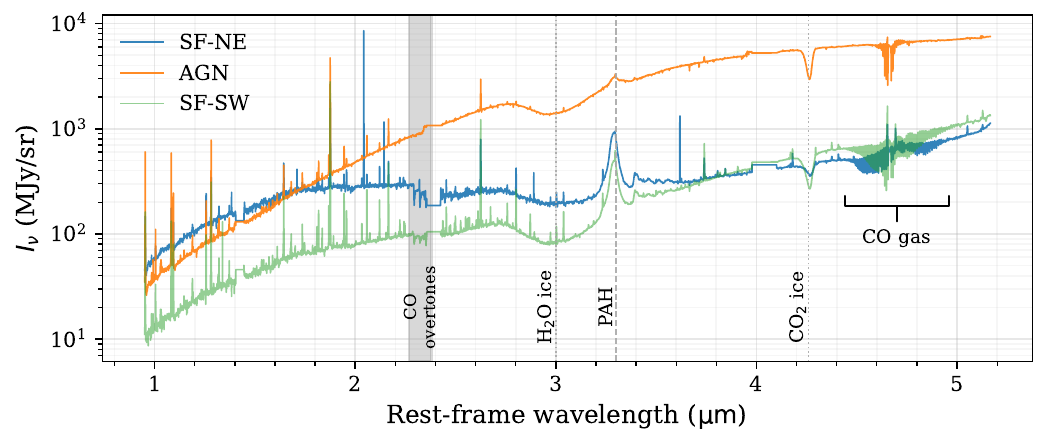}
    \caption{NIRSpec spectra towards the three nuclear regions under consideration. 
    The spectral regions close to \SI{1.4}{\micron}, \SI{2.4}{\micron} and \SI{4.05}{\micron} rest-frame wavelengths are affected by the NIRSpec detector gaps; {these parts of the spectrum are masked out in this figure. Several prominent features are indicated. The AGN spectrum rises much more steeply between \SI{1}{\micron} and \SI{4}{\micron} than the spectra of the star-forming regions.} 
    }
    \label{fig:nirspec_spectra_full}
\end{figure*}

\begin{figure*}[ht!]
    \centering
    \includegraphics[width=\linewidth]{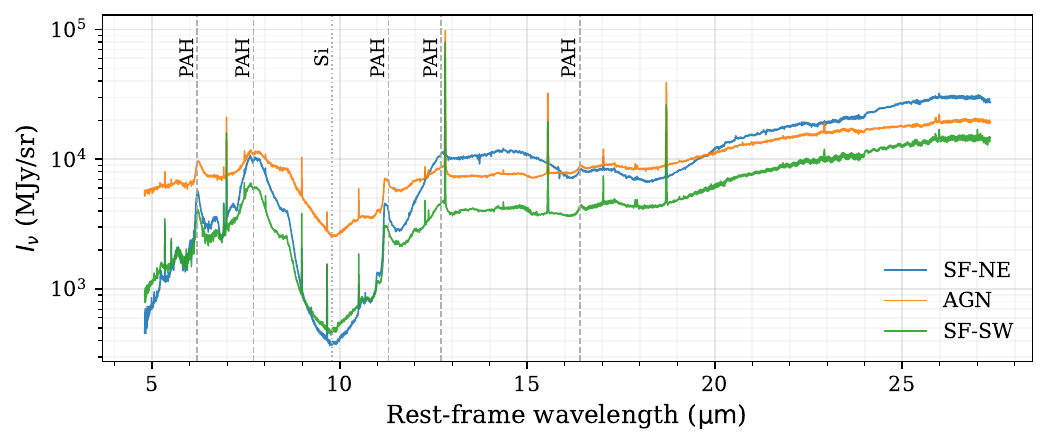}
    \caption{MIRI MRS spectra towards the three regions under consideration. {Prominent PAH emission features and the deep \SI{9.7}{\micron} silicate absorption band are labeled. The AGN spectrum is relatively flat and exhibits only a moderately deep \SI{9.7}{\micron} silicate absorption band compared to the spectra of the starburst regions.}
    }
    \label{fig:mrs_spectra_full}
\end{figure*}

We present the 1D spectra extracted from the three apertures over the full wavelength range of the instruments in Fig.~\ref{fig:nirspec_spectra_full} (NIRSpec) and Fig.~\ref{fig:mrs_spectra_full} (MIRI MRS). Here we briefly discuss the continuum shape and the most prominent features to contextualise our analysis.

The overall view of the spectra 
immediately reveals several striking differences between the AGN position on the one hand and the SF-NE and SF-SW knots on the other hand. The AGN spectrum shows a steeply rising continuum from 1 to \SI{4}{\micron}, which remains relatively flat at longer wavelengths. This behaviour is characteristic of dust-embedded AGNs, and typically attributed to hot dust emission directly associated with the circumnuclear torus \citep[e.g.,][]{Barvainis1987,PierKrolik1992,HonigKishimoto2010,GonzalezMartin2019}.
The SF-NE and SF-SW spectra, on the other hand, exhibit very similar continuum shapes, which are flatter than the AGN spectrum between 1 and \SI{4}{\micron} but have a stronger rise at wavelengths above about \SI{18}{\micron}.
The $\SI{9.7}{\micron}$ silicate absorption feature is very deep in the two starburst regions, 
indicating deeply dust-embedded star formation;
the silicate feature is considerably shallower at the AGN position. Furthermore, the PAH emission features at $\SI{3.3}{\micron}$, $\SI{6.2}{\micron}$, $\SI{7.7}{\micron}$ and $\SI{11.3}{\micron}$ are strong towards the SF-NE and SF-SW regions, but nearly vanish in the AGN spectrum \citep[see also][]{Rich2023}. These general properties{, which are in agreement with the ideas of \citet{Laurent2000} and \citet{Marshall2018},} all suggest the presence of a moderately obscured AGN at the position ``AGN'', and more deeply embedded star formation at positions SF-NE and SF-SW\null.

Several broad ice absorption features are detected as well, {towards all three regions under consideration}. At $\SI{3}{\micron}$, the broad \ce{H2O} stretch \citep[e.g.][]{Boogert2015} is observed. The \ce{CO2} stretch is seen at $\SI{4.27}{\micron}$; interestingly, it is weakest in the most deeply obscured star-forming region (SF-NE). None of the spectra show considerable CO ice absorption at $\SI{4.7}{\micron}$, but all show strong CO gas-phase absorption at this wavelength.

\subsection{Fundamental CO band}

\begin{figure*}[ht!]
    \centering
    \includegraphics[width=\linewidth]{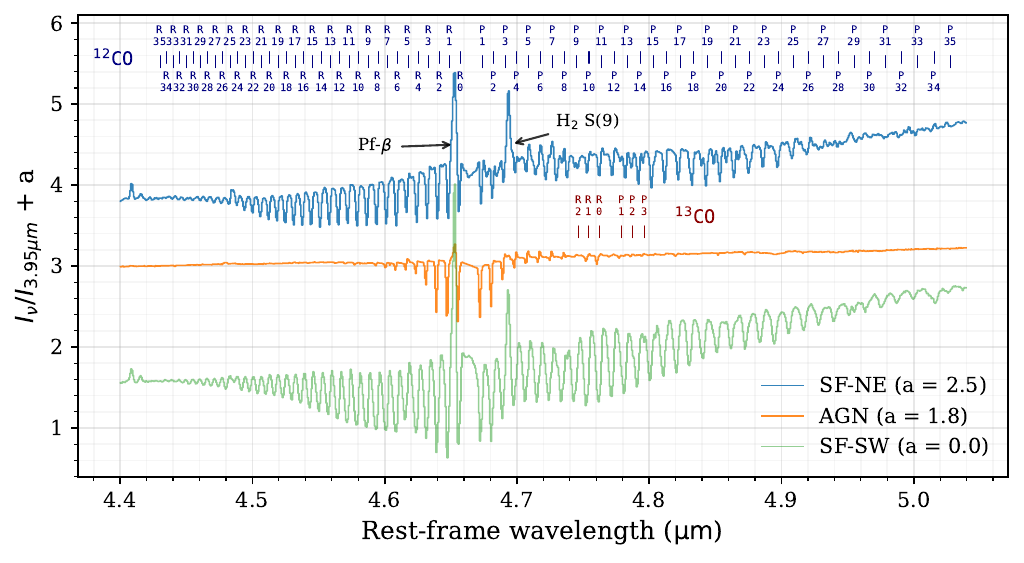}
    \caption{The fundamental CO band as observed towards three nuclear regions in VV\,114E. The intensities are offset by a factor $a$ for visibility (see legend). The lines of the $^{12}$CO fundamental band are labeled in dark blue; some $^{13}$CO lines are indicated in red. Towards the SF-NE and SF-SW regions we see broad, deep bands, indicating that the molecular gas is highly excited. The AGN position exhibits a narrow $^{12}$CO band; the R(1), R(0) and P(1) lines of $^{13}$CO are also detected here.}
    \label{fig:CO_5micron_spec}
\end{figure*}

The fundamental band of CO, arising from absorption from the vibrational ground state ($v=0$) to the first excited state ($v=1$), appears prominently as a sequence of deep lines in all three regions. These bands are shown in detail in Fig.~\ref{fig:CO_5micron_spec}. The CO bands are comprised of an R-branch bluewards of $\SI{4.67}{\micron}$, corresponding to transitions with $\Delta J = +1$ in absorption, and a P-branch on the red side, where transitions satisfy $\Delta J = -1$. Here $\Delta J = J_{\mathrm u} - J_{\mathrm l}$, where $J_{\mathrm u}$ and $J_{\mathrm l}$ are the rotational quantum numbers of the upper and lower level, respectively. At the SF-NE and SF-SW positions, we observe broad, deep bands, indicating highly-excited CO gas against a bright background continuum. The AGN region, however, only exhibits a narrow absorption band, with some weak emission present in the form of P-Cygni profiles in the P-branch.

The R-branch is contaminated by two weak 
{emission} lines, partially filling in the R(5) and R(24) lines. Furthermore, the Pf-$\beta$ emission line at $\SI{4.654}{\micron}$ calls for some care in using the R(0) and R(1) lines. The P-branch suffers from more contaminants: the \ce{H2} 0-0 S(9) emission line lies at $\SI{4.695}{\micron}$, but the fundamental bands of $^{13}$CO and C$^{18}$O run through the $^{12}$CO P-branch as well. 

\subsection{H$_2$O}

\begin{figure*}
    \centering
    \includegraphics[width=\linewidth]{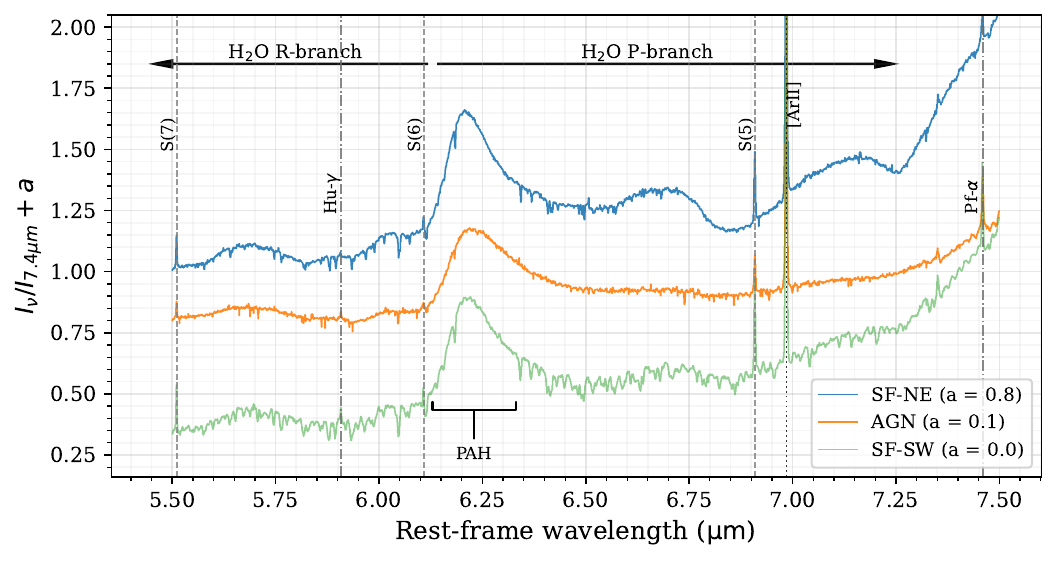}
    \caption{Water vapour absorption lines in the MIRI MRS spectra. They are most prominent at the SF-SW position. The location of the \ce{H2O} R- and P-branches are indicated by black arrows. Several emission lines are marked: dashed lines indicate the \ce{H2} S(5), S(6) and S(7) lines, a dotted line indicates the [ArII] line, and dash-dotted lines indicate hydrogen recombination lines. The intensities are offset by a factor $a$ for visualisation purposes.}
    \label{fig:h2o-spectra}
\end{figure*}

We detect rovibrational lines of water vapour between approximately 5.5 and \SI{7.2}{\micron} towards all three regions under consideration; they are shown in detail in Fig.~\ref{fig:h2o-spectra}. The H$_2$O lines are weak and narrow towards the AGN and more clearly visible towards the SF-NE and SF-SW regions, where they are significantly broader and deeper.

This spectral region also encompasses the strong $\SI{6.2}{\micron}$ PAH feature, as well as prominent absorptions at $\SI{6.85}{\micron}$ and $\SI{7.25}{\micron}$. We detect the latter only in the deeply-embedded SF-NE core. These features are typically attributed to aliphatic or hydrogenated amorphous hydrocarbons \citep{Spoon2001,Spoon2004,Rich2023}; for further discussion of these assignments see \citet{Boogert2015}.

\subsection{\ce{C2H2} \& \ce{HCN}}

\begin{figure*}[ht!]
    \centering
    \includegraphics[width=\linewidth]{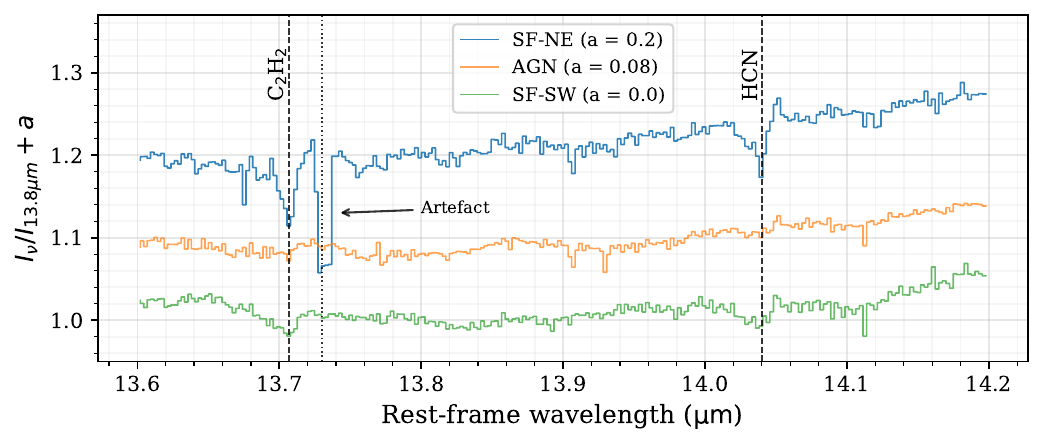}
    \caption{MIRI MRS spectra at $\sim \SI{14}{\micron}$. The spectra for the different apertures are offset by a term $a$ (see legend) for visibility. No 1D residual fringe correction was applied here. The Q-branches of HCN and \ce{C2H2} are clearly detected towards the SF-NE region, and are tentatively detected towards the SF-SW region as well.}
    \label{fig:spectra-14micron}
\end{figure*}

At longer wavelengths, our spectral and spatial resolution decrease, impeding the detection of shallow individual lines. Fig.~\ref{fig:spectra-14micron} shows the three spectra between $\SI{13.6}{\micron}$ and $\SI{14.2}{\micron}$. As reported previously by \cite{Rich2023}, we detect the blended Q-branches of both \ce{C2H2} and HCN at $\SI{13.7}{\micron}$ and \SI{14.04}{\micron} at the SF-NE position. They may be present in the SF-SW spectrum as well, but the signal-to-noise ratio here is too low to confirm a detection.

{The abundances of both HCN and \ce{C2H2} can be greatly enhanced by high-temperature chemistry in the inner envelopes around young massive stars \citep[e.g.][]{Doty2002, Rodgers2003}.} HCN is of {particular} interest as its rotational transitions in the $\nu_2 = 1$ state have been detected towards several U/LIRGs \citep[e.g.][]{Sakamoto2010,Aalto2015a,Aalto2015b,Falstad2021}. With MIRI MRS, we observe the absorption due to infrared pumping by the strong $\si{14}{\micron}$ continuum, causing the excitation to the $\nu_2 = 1$ state. {For \ce{C2H2}, which has no dipole-allowed rotational transitions, mid-IR observations are a unique detection method.}

\subsection{CO overtones} \label{subsec:co_overtones}
\begin{figure}
    \centering
    \includegraphics[width=\linewidth]{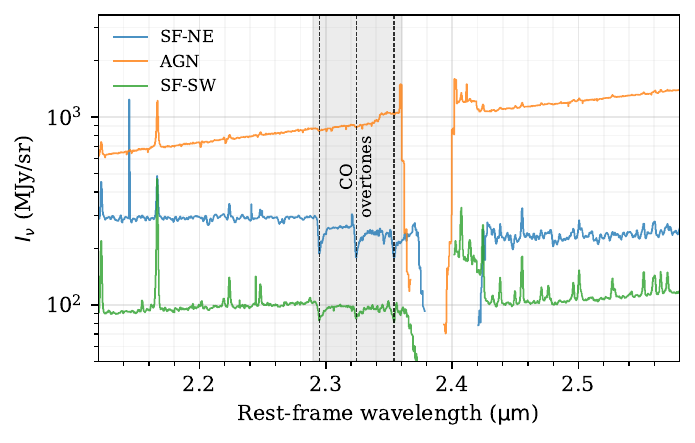}
    \caption{CO overtone bands at 2.3 micron. From left to right, the dashed lines indicate the $v = 0-2$ band, the $v = 1-3$ band, and the $v = 2-4$ band; higher bands could not be observed due to the detector gap. The CO absorption features are detected towards the SF-NE and SF-SW positions, but not towards the AGN position.}
    \label{fig:co_overtones}
\end{figure}

In addition to the fundamental CO band from the $v = 0-1$ transitions at $\SI{4.7}{\micron}$, the 1st overtones{---rovibrational transitions for which $\Delta v = \pm 2$---} at $\SI{2.3}{\micron}$ lie in the NIRSpec spectral range as well. This part of the spectrum is shown in Fig.~\ref{fig:co_overtones}. We detected these absorption bands towards the SF-NE and SF-SW positions, but not towards the AGN position.
The bands display prominent bandheads at the blue side, a shape that is characteristic of the spectra of the atmospheres of cool stars, such as red giants and supergiants. {The latter are ubiquitous in starburst regions \citep[e.g.,][]{Oliva1995,Armus1995}. The $\SI{2.3}{\micron}$ CO bands appear a few Myr after the onset of star formation, and remain prominent for more than $\SI{e8}{years}$ \citep{Origlia1999,Leitherer1999,OrigliaOliva2000}.}
The lack of these CO bands towards the AGN position therefore unambiguously confirms its nature as an AGN\null.


\section{Analysis}
\subsection{CO} \label{subsec:analysis-co}
The CO band spectra presented in Fig.~\ref{fig:CO_5micron_spec} display a remarkable dichotomy. The SF-NE and SF-SW positions show broad vibrational bands, revealing high rotational excitation temperatures. In contrast, the narrow vibrational band observed at the AGN position indicates a much lower rotational excitation. CO rotational ladders observed {at lower spatial resolution} in the (sub)millimeter and far-infrared, however, show exactly the opposite behaviour, with the highest rotational excitation observed for AGN, and lower rotational excitation for starbursts \citep{vdWerf2010,Rosenberg2015,Lu2017}. As discussed previously, the star formation nature of SF-NE and SF-SW, and the AGN nature of the so labeled position are supported by strong evidence from infrared and radio observations \citep{Evans2022, Rich2023}. The broad CO bands at the two SF positions and the narrow band at the AGN position are therefore very surprising. 

Determining the cause of this behaviour is the focus of the present paper.
To proceed, we first analyse the various molecular bands using rotational population diagrams, in order to quantify the excitation. We then discuss the physical origin of these results, the underlying physics, and the implications in Section~\ref{sec:discussion}.

We begin by constructing rotational population diagrams based on our three fundamental CO band spectra. The principle has been discussed in the literature before, but usually in treatments tailored to rotational line emission \citep[e.g.,][]{GoldsmithLanger1999}. Since here we will be deriving rotational excitation temperatures from vibrational absorption lines, we provide the relevant equations, and details of our methods, in Appendix~\ref{appendix:rot_diagrams}; a summary is provided here.

We treat the spectra as pure absorption and convert them to optical depth spectra through $I_\text{obs} = I_\text{cont} e^{-\tau}$, using a spline fit to
spectral regions that appear to be line-free to estimate the local continuum $I_\text{cont}$. We then fit a Gaussian profile to each line in optical depth space and take the integral of this model to obtain an integrated optical depth. Using
Einstein $A$ values from HITRAN\footnote{\url{https://hitran.iao.ru}} (\citealp{Rothman2013}, with values for CO from  \citealp{Guelachvili1983, Farrenq1991, Goorvitch1994}), we calculate the lower-level column density for each detected line {using Eq. \ref{eq:coldens_practical}}.

The distribution of these vibrational ground state level populations allows us to probe the physical conditions of the CO gas by comparing to the Boltzmann distribution. We employ a rotation diagram analysis to characterise the {rotational excitation temperature $T_\mathrm{ex}$, defined as:

\begin{equation} \label{eq:T_ex}
    \frac{N_i}{N_j} \equiv \frac{g_i}{g_j} e^{-(E_i - E_j)/(k T_\mathrm{ex})}
\end{equation}

Here $N_i$, $g_i$ and $E_i$ denote the column density, statistical weight, and energy of level $i$; $k$ is the Boltzmann constant.
In the rotation diagram, we plot the 
logarithm of $N_i/g_i$ as a function of the level energy in temperature units. The slope is then governed by the excitation temperature of the gas
(see Appendix \ref{appendix:rot_diagrams}).



The CO rotation diagrams for the three regions under consideration are shown in Fig.~\ref{fig:co_rot_diagram} (SF-NE \& SF-SW) and \ref{fig:co_rot_diagram_sw} (AGN). For the rotation diagram analysis, we only use the R-branch lines, as some P-branch lines appear to be affected by emission and other contaminants (Fig.~\ref{fig:CO_5micron_spec}). We also remove the R(5) and R(24) lines from the fits, as these lines are contaminated by faint emission from other species.

\begin{figure*}[ht!]
    \centering
    \includegraphics[scale=0.73]{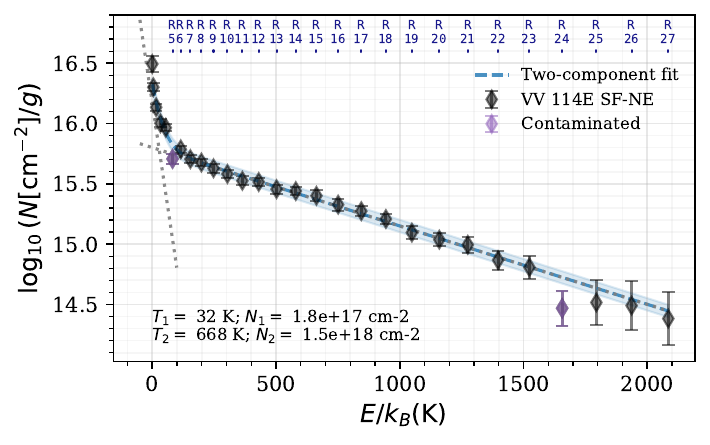}
    \includegraphics[scale=0.73]{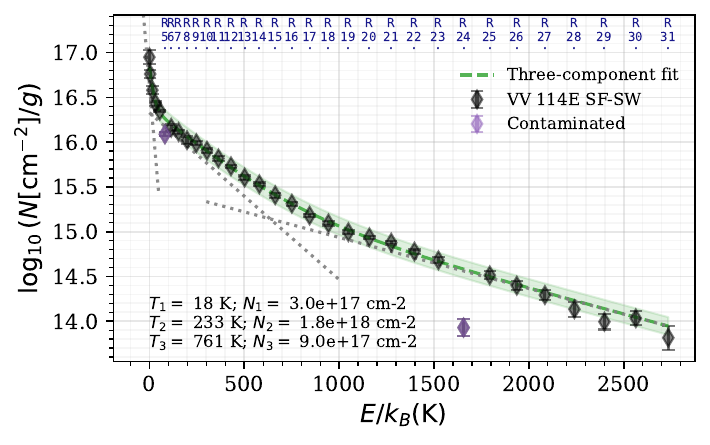}
    \caption{Rotation diagrams from the R-branch lines of CO towards the star-forming cores, including a two-component (SF-NE)/three-component (SF-SW) LTE model fit to the points. {Grey dotted lines indicate the LTE models for each individual component.} In both regions, the R(5) and R(24) lines are partially filled in by weak emission from other species, leading to an underestimation of the CO level column density; the corresponding points are marked in purple on the rotation diagram and were left out of the fit.}
    \label{fig:co_rot_diagram}
\end{figure*}

\begin{figure}[ht!]
    \centering
    \includegraphics[scale=0.73]{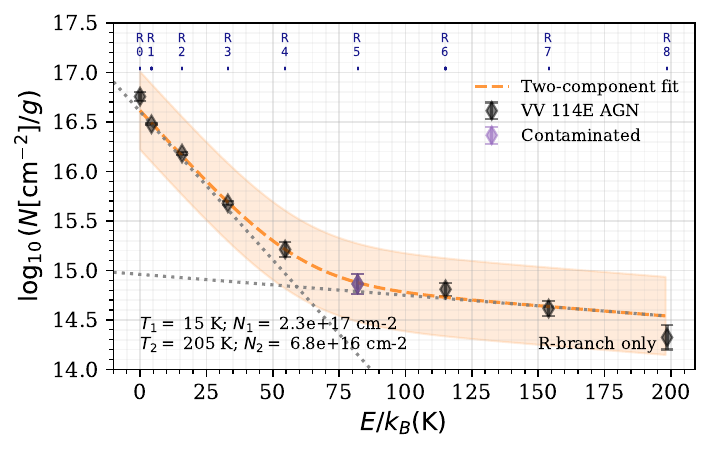}
    \caption{Rotation diagram from the R-branch lines of CO towards the AGN position. {Grey dotted lines indicate the LTE models for each individual component.} The $J \leq 3$ levels are likely dominated by cool foreground gas.}
    \label{fig:co_rot_diagram_sw}
\end{figure}

At the SF-NE position, the CO rotation diagram exhibits a low-excitation component in the first 3-5 $J$-levels, followed by an extended tail of seemingly thermalised gas that continues out to at least $J = 23$. To quantify the excitation, we fit a model combining two LTE components (Eq.~\ref{eq:rot_diagram_LTE}) to the measurements. The high-excitation tail is well-fit by a rotational temperature $T_\mathrm{ex} \approx \SI{700}{\kelvin}$.


For the SF-SW region, the CO rotation diagram similarly shows a low-excitation component up to $J = 4$ and an extended high-excitation tail. However, here the tail has another kink at ${E_l/k_B \sim \SI{800}{\kelvin}}$, indicating two distinct high-temperature components. Fitting a three-component LTE model, we find rotational temperatures of $T_\mathrm{ex} \approx {\SI{200}{\kelvin}}$ and ${T_\mathrm{ex} \approx \SI{800}{\kelvin}}$ for the $J > 5$ levels.


At the AGN position, the picture is quite different: the rotation diagram is well-fit by two relatively cool components. {We do not detect a high-excitation component. Given the sensitivity of these current data, we exclude the presence of a $T_\mathrm{ex} \approx \SI{700}{\kelvin}$ component with column density $N(\mathrm{CO}) \gtrsim \SI{3e16}{\centi\meter^{-2}}$ along the line of sight---two orders of magnitude lower than the column densities inferred for the SF-NE and SF-SW regions.}

Resulting parameters for all fits are displayed in the figures and summarized in Table~\ref{tab:results_summary}. All positions show a relatively cool (${T_\mathrm{ex}=15-\SI{32}{K}}$) gas component dominating the lowest rotational levels. In all likelihood, this component corresponds to the bulk molecular gas in the system \citep{Saito2015}, probably subthermally excited at the SF-NE and SF-SW positions. This gas is probably ``foreground'' and not directly associated with the nuclear mid-IR cores.
The fitting results confirm the presence of highly-excited CO gas in two presumed star-forming regions and absence thereof at the AGN position.
We will return to this important issue in Section \ref{sec:discussion_AGN_SB}.
\\


We extract the velocity dispersion and radial velocity with respect to the velocity of warm \ce{H2} from the optical depth line profile fit for each of the lines. We remove the contaminated R(5) and R(24) lines from this analysis. The measured velocity dispersions are corrected for the instrumental broadening, assuming the nominal resolving power of $R = 2700$ \citep{Jakobsen2022}. The results are shown in Fig.~\ref{fig:co-velocities}.

\begin{figure}[ht!]
    \centering
    \includegraphics[width=\linewidth]{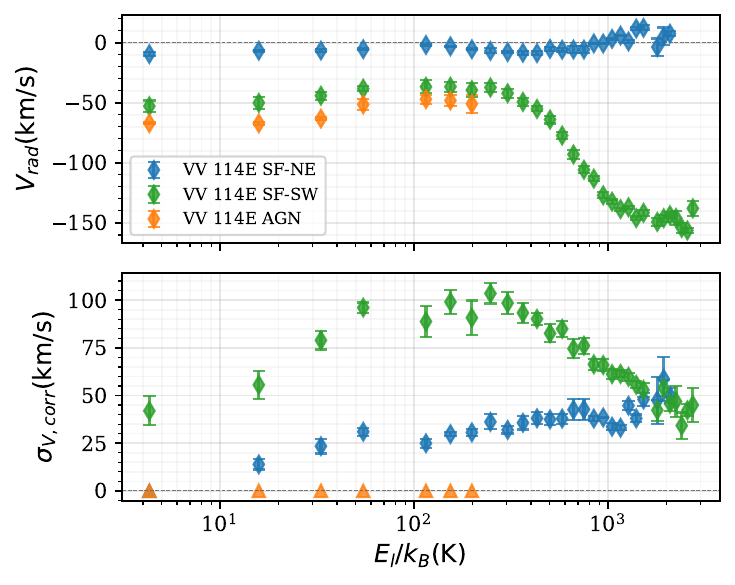}
    \caption{Radial velocities (top) and instrument-corrected velocity dispersions (bottom) of the CO R-branch lines as a function of the energy of the lower level involved in the transition. Grey dashed lines indicate the $\SI{0}{\kilo\meter\per\second}$ line w.r.t. \ce{H2} emission lines. {Towards the AGN, the CO lines are unresolved; these are indicated by triangles in the bottom panel.}
    }
    \label{fig:co-velocities}
\end{figure}

Fig.~\ref{fig:co-velocities} shows that the CO gas observed towards SF-NE is effectively at the velocity of the \ce{H2} lines throughout the band. Towards the AGN, however, the CO exhibits a clear blueshift of $V_\mathrm{rad} \approx \SI{-60}{\kilo\meter\per\second}$. At the SF-SW position, the radial velocity transitions from $V_\mathrm{rad} \approx \SI{-50}{\kilo\meter\per\second}$ at the low-excitation CO lines to $V_\mathrm{rad} \approx \SI{-140}{\kilo\meter\per\second}$ at the highest $J$-lines, suggesting the presence of slightly blueshifted cool gas and more rapidly outflowing 
highly-excited gas.

{{Towards the star-forming regions}, the velocity dispersions reveal a distinct lower-dispersion component that corresponds to the cold component seen in the rotation diagrams (Fig.~\ref{fig:co_rot_diagram} and \ref{fig:co_rot_diagram_sw}). The SF-SW region exhibits a peak in velocity dispersion of $\sigma_V \approx \SI{100}{\kilo\meter\per\second}$ at the R(4)-R(7) lines---where the rotation diagram transitions from the foreground component to the $T_\mathrm{ex} \approx \SI{230}{\kelvin}$ component, and the radial velocities start to transition to $V_\mathrm{rad} \approx \SI{-140}{\kilo\meter\per\second}$. The peak in measured velocity dispersion is therefore explained by the blending of two lines arising from CO gas at two distinct velocities, {with velocity dispersions of $\sim \SI{40}{\kilo\meter\per\second}$ for the individual kinematic components. These estimates are roughly consistent with dispersions found for rotational emission lines \citep{Saito2015, Saito2018}.} Although the rotation diagram is indicative of three temperature components, we find no evidence of a third velocity component.}
\\

We model the three CO spectra using the column densities and excitation temperatures of the various components inferred from the rotation diagram, and the velocity dispersions of the lines dominated by specific components, as collected in Table \ref{tab:results_summary}. These models are shown 
{together with} the continuum-extracted observed spectra in Fig.~\ref{fig:co_model_spec_sw}. {The models spectra of the individual components are shifted by the measured radial velocities listed in Table \ref{tab:results_summary}.}

For the star-forming regions, the correspondence between data and model is good in the R-branch---from which the rotation diagram was constructed---but less so in the P-branch. Towards the SF-NE core, the observed P-branch absorption lines are considerably shallower than predicted by the model. 
This asymmetry has significant implications, which we shall discuss in Section \ref{sec:discussion-excitation}. {We note that imperfections in the SF-SW model are most likely due to the blending of at least two kinematic components that were not explicitly separated in the construction of the rotation diagram.}

In the AGN region, the observed absorption lines match the model well in both branches, but they are blueshifted with a radial velocity of $-\SI{55}{\kilo\meter\per\second}$. Most notably, the P(4), P(5), P(6), P(7) and P(8) lines clearly appear in emission as well, at the systemic velocity. These P-Cygni profiles suggest the presence of a molecular outflow. At the SF-SW position, just southeast of the AGN, the high-excitation lines are blueshifted at a radial velocity up to $-\SI{140}{\kilo\meter\per\second}$, while the low-excitation lines are shifted by only $-\SI{50}{\kilo\meter\per\second}$.

\begin{figure}[ht!]
    \centering
    \includegraphics[width=\linewidth]{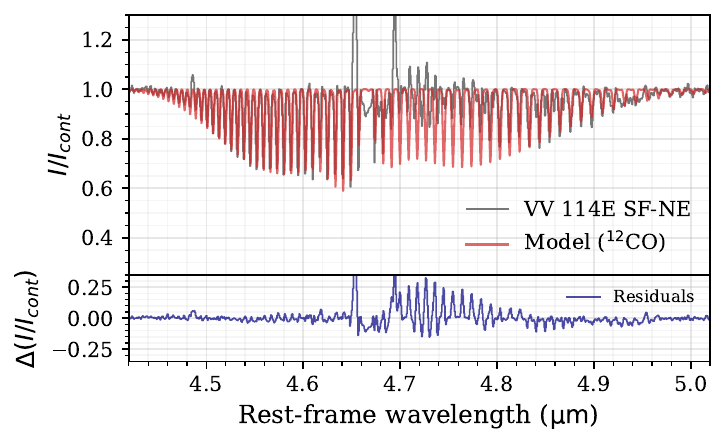}
    \includegraphics[width=\linewidth]{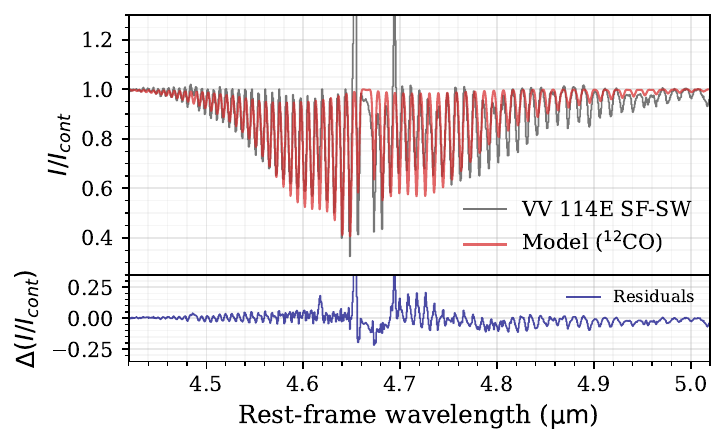}
    \includegraphics[width=\linewidth]{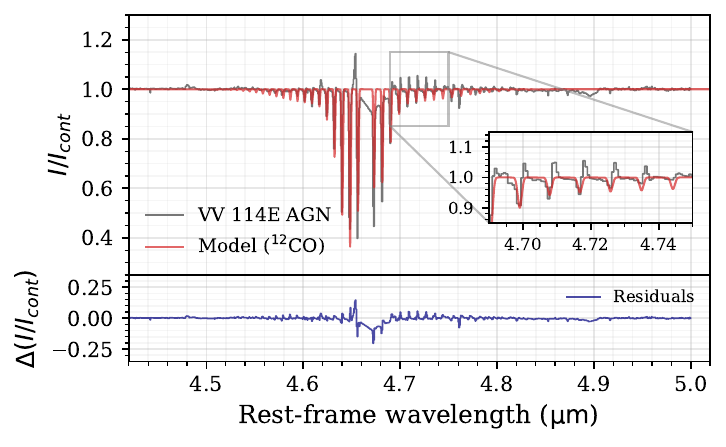}
    \caption{The continuum-extracted observed CO spectra vs. the model spectra corresponding to the best-fit two-component (SF-NE, AGN) or three-component (SF-SW) LTE model. 
    {The model spectra are shifted to match the measured radial velocities. The model-subtracted spectra are presented as well.}
    The AGN spectrum (bottom panel) is shown on a smaller spectral scale to highlight the blueshift of the absorption lines; an enlarged view of the P-branch P-Cygni profiles is included. The weak, broad feature at $\approx\SI{4.68}{\micron}$ is due to CO ice.
    }
    \label{fig:co_model_spec_sw}
\end{figure}

{The above analysis makes the implicit assumption that the absorbing gas fully covers the background continuum source. If this is not the case, our calculation underestimates optical depths and column densities. The derived temperatures are, however, not strongly affected if the peak optical depths of the high excitation lines remain small compared to 1, since in this case the column density ratios remain unaltered. High optical depths would lead to a situation where the strongest lines saturate and have approximately the same depth. Inspection of Figs.~\ref{fig:CO_5micron_spec} and \ref{fig:co_model_spec_sw} reveals that this is clearly not the case at the AGN position and at position SF-SW\null. However, at position SF-NE a number of the deepest lines indeed have similar depths, and the possibility of saturation must be considered.} 

To assess the effect that incomplete coverage of the background continuum might have on the inferred parameters for the SF-NE region, we assume the minimum covering factor {of $f_c = 0.4$---a lower covering factor would not be able to produce the apparent absorption depth of 40\% that the deepest line observed here reaches}---and correct the observed absorption lines for this covering factor. We then construct a rotation diagram from this corrected spectrum. The implied column densities increase by a factor of approximately 3.5 compared to the case where we assume $f_c = 1$; the inferred temperature of the high-excitation component decreases by {19}\% to ${T_\mathrm{ex} = \SI{559 \pm 16}{\kelvin}}$. {A similar exercise for the SF-SW region with covering factor $f_c = 0.7$ leads to a column density increase of a factor 1.5 and a temperature decrease of 6\% to $T_\mathrm{ex} = \SI{717 \pm 48}{\kelvin}$.} Thus, even in the most extreme possible case, the high excitation temperature measured is not strongly diminished due to saturation of the lower$-J$ lines.

{{The CO column density corresponds to a $\SI{4.5}{\micron}$ dust optical depth $\tau_{4.5}$ which can be calculated from:} 

\begin{equation}
    {\tau_{4.5}=\frac{1}{1.086}\,\frac{\tau_{4.5}}{\tau_V}\,\frac{A_V}{N(\mathrm{H})}\,\frac{\mathrm{2[H_2]}}{\mathrm{[CO]}}\,N(\mathrm{CO})}
\end{equation}

Using a standard molecular cloud CO abundance $\mathrm{[CO]/[H}_2]=10^{-4}$, the Milky Way infrared extinction curve determined by \citet{Schlafly2016}, and $A_V/N(\mathrm{H})=\SI{5e{-22}}{\mathrm{mag}\,\centi\meter\squared\,\per\mathrm{H}}$ \citep{Bohlin1978,Rachford2009}, we find that $\tau_{4.5}=1$ corresponds to $N(\mathrm{CO})=\SI{3.3e18}{\per\centi\meter\squared}$. The dust optical depths implied by the CO column densities in Table~\ref{tab:results_summary} are then $\tau_{4.5}=0.5$ for SF-NE and $\tau_{4.5}=0.9$ for SF-SW\null. Applying the proposed increase of the column density by a factor 3.5 in SF-NE would result in $\tau_{4.5}\approx1.8$ for that region. This number exceeds the value derived from the depth of the $\SI{9.7}{\micron}$ silicate absorption feature \citep[their Figs.~4 and A1]{Donnan2023} by less than a factor of 2. With the corrected column density for SF-NE, the relative depths of the $\SI{9.7}{\micron}$ silicate feature for SF-NE and SF-SW are also reproduced.}



\subsection{\ce{H2O}} \label{sec:analysis-h2o}

For water, constructing a rotation diagram is more difficult than for CO as many apparent lines are actually blends of several lines. In particular for the SF-SW region, where the \ce{H2O} lines are considerably broader, many lines had to be removed from the 
{fit}. {Modeling these additional lines would require a more elaborate procedure \citep[see][]{Gonzalez-Alfonso2023}, which is not necessary for the purposes of the present paper.} 
The R-branch \ce{H2O} rotation diagrams for the two starburst regions are presented in Fig.~\ref{fig:rot_diagrams_h2o}. 
We fit a single-component LTE model to the data. We do not attempt a fit for the weak \ce{H2O} lines at the AGN position.

\begin{figure*}[ht!]
    \centering
    \includegraphics[scale=0.73]{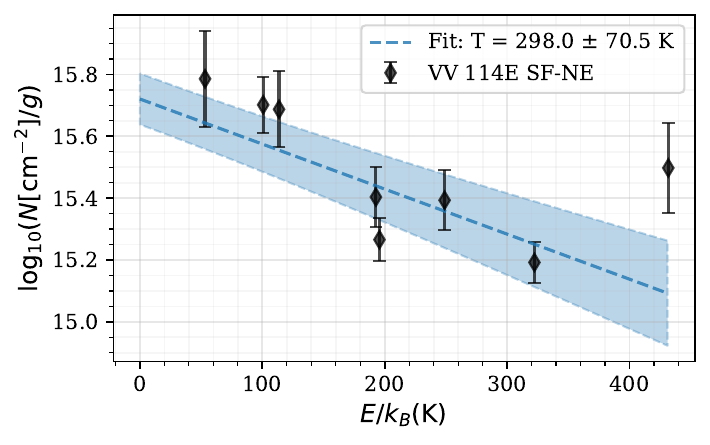}
    \includegraphics[scale=0.73]{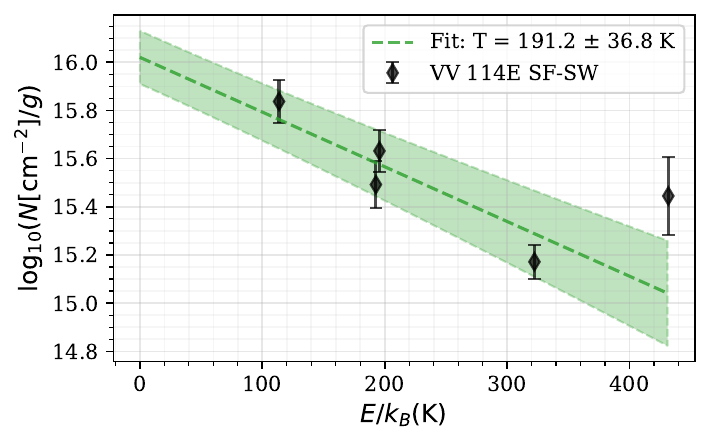}
    \caption{Rotation diagrams of water vapour R-branch lines towards the SF-NE and SF-SW positions. The single-component LTE fit and its inferred excitation temperature is indicated.}
    \label{fig:rot_diagrams_h2o}
\end{figure*}

For the lines selected for the rotation diagram analysis, we measure the radial velocities and velocity dispersions as described in Section \ref{subsec:analysis-co}, now assuming an instrumental resolving power $R = 3400$ \citep{Labiano2021}. The results, 
summarised in Table \ref{tab:results_summary}, again show gas observed towards the SF-NE core that is effectively at systemic velocity, while the \ce{H2O} absorption at the SF-SW position is considerably blueshifted. In this region, the radial velocity of the water lines matches that of the highly-excited CO lines, with $V_\mathrm{rad} \approx \SI{-110}{\kilo\meter\per\second}$. Interestingly, in both cores the velocity dispersion of the \ce{H2O} lines is significantly higher than those of CO.


Using the rotation diagram results obtained from the R-branch lines, and the median velocity dispersions, we model the water spectra for the SF-NE and SF-SW regions with a single-component LTE model; they are presented 
{together with} the observed spectra in Fig.~\ref{fig:h2o-model-spec}. At both positions, the match is reasonable in the R-branch, but we see considerably shallower absorption than predicted in the P-branch. The same P-R branch asymmetry was noted for CO (Section \ref{subsec:analysis-co}). We discuss the origin of this asymmetry between branches in Section \ref{sec:discussion-excitation}.

\begin{figure*}[ht!]
    \centering
    \includegraphics[width=\linewidth]{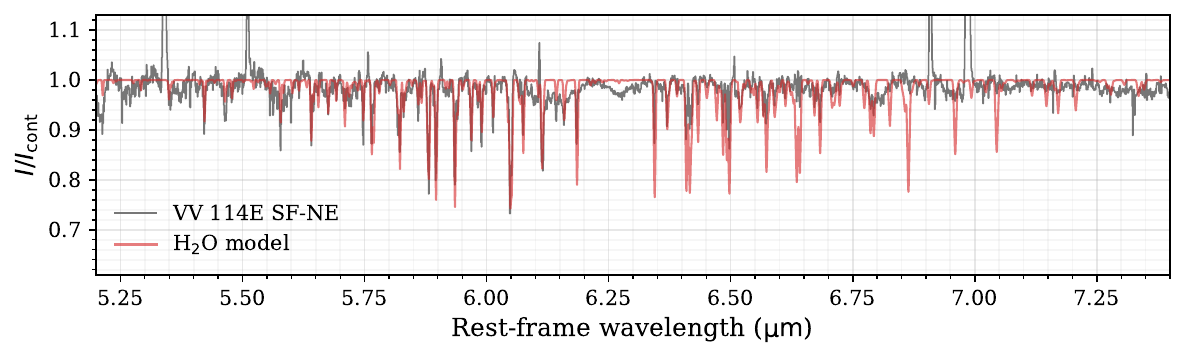}
    \includegraphics[width=\linewidth]{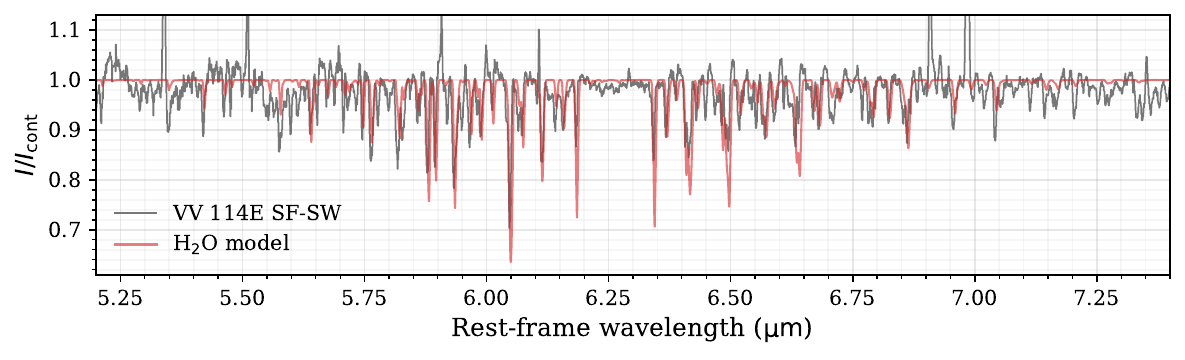}
    \caption{The continuum-extracted \ce{H2O} spectra observed towards the NE and SW-E cores vs.\ the LTE model spectra derived from the rotation diagrams in Fig.~\ref{fig:rot_diagrams_h2o}. The models match the observed spectra quite well, taking into account errors in the continuum fits.}
    \label{fig:h2o-model-spec}
\end{figure*}
\subsection{\ce{C2H2} \& HCN} \label{subsec:analysis-14micron}
For the strong Q-branches of \ce{C2H2} and HCN at $\sim \SI{14}{\micron}$, we cannot use the rotation diagram method because we do not resolve the individual lines. Instead, we use a direct spectral fitting analysis, employing a grid of single-component LTE models and $\chi^2$ minimisation. For both molecules, we use a $30 \times 50$ grid with column density spanning a range $\log_{10} (N/\SI{}{\centi\meter^{-2}}) \in (15.0, 16.5)$ and temperature spanning $T/\SI{}{\kelvin} \in (50, 500)$. The column densities are logarithmically distributed. We fix the velocity dispersion of the individual lines at $\sigma_V = \SI{50}{\kilo\meter\per\second}$---the value we found for the water lines---but through experimentation we find that this value has little effect on the inferred parameters.
We limit the spectral range used in the fits to the fundamental Q-branch, and we only model the strong \ce{C2H2} and HCN lines at the SF-NE position.

To account for the noise in the spectrum, we apply a bootstrap procedure: we resample the spectral pixels used in the fit 5000 times, and take the best-fit model from each iteration. The resulting corner plots are shown in Fig.~\ref{fig:bootstrap_corner_HCN_C2H2}. To visualise the spread in the best-fit models, we draw 50 random bootstrap samples and plot the corresponding models along with the observed spectrum---these fits are shown in Fig.~\ref{fig:bootstrap_spectra_HCN_C2H2}. Fit results are again displayed in the figures and in Table \ref{tab:results_summary}.

\begin{figure*}[ht!]
    \centering
    \includegraphics[scale=0.65]{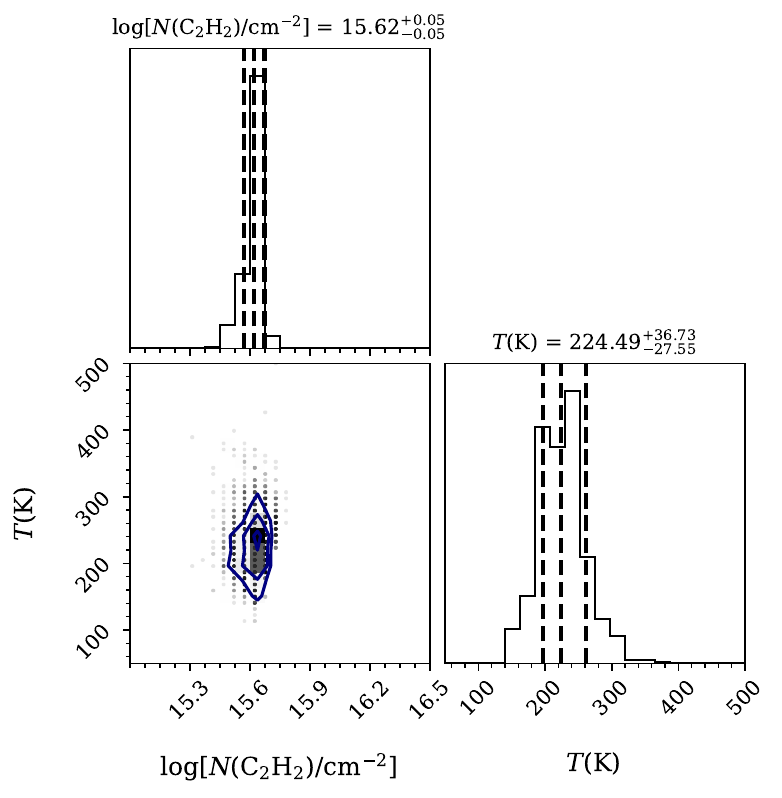}
    \includegraphics[scale=0.65]{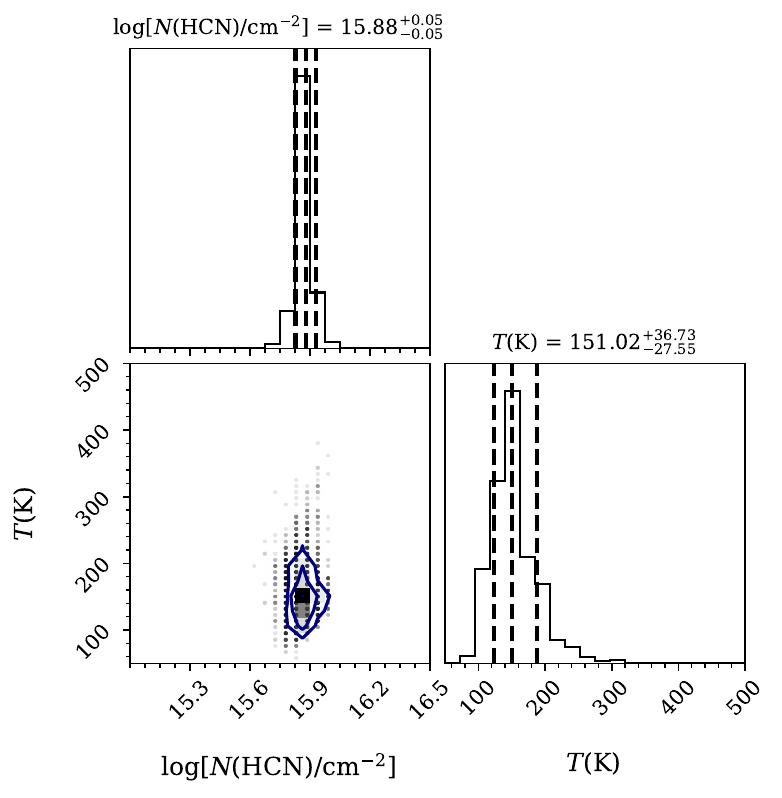}
    \caption{Visualisation of the distribution of bootstrap samples for \ce{C2H2} (left) and HCN (right) towards the SF-NE core. The velocity dispersion of individual lines was fixed at $\SI{50}{\kilo\meter\per\second}$, but this exact value does not significantly affect the band shape. Dashed lines indicate 68\% confidence intervals on the parameters. The column densities and excitation temperatures are reasonably well-constrained, but the distributions are somewhat asymmetrical.}
    \label{fig:bootstrap_corner_HCN_C2H2}
\end{figure*}

\begin{figure*}[ht!]
    \centering
    \includegraphics[scale=0.72]{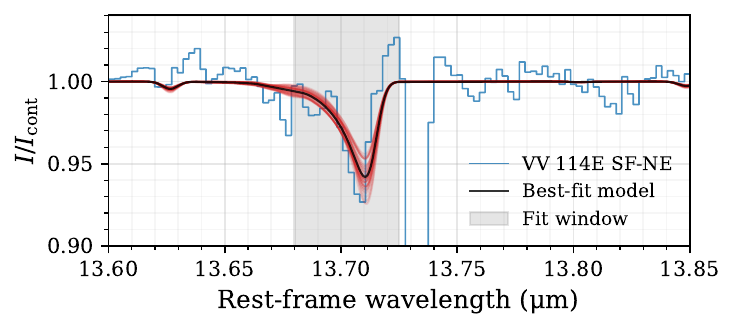}
    \includegraphics[scale=0.72]{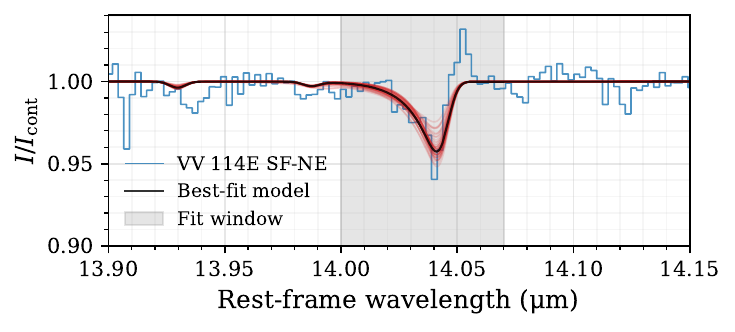}
    \caption{Model spectra for \ce{C2H2} (left) and HCN (right), based on the grid bootstrap fits. The red curves represent 50 random samples from the bootstrap; the shaded grey area indicates the spectral region used in the fit. The sampled models are reasonably good fits, but exhibit significant scatter.}
    \label{fig:bootstrap_spectra_HCN_C2H2}
\end{figure*}

\begin{table*}[ht!]
    \centering
    \begin{tabular}{cccccccc}
    \multicolumn{8}{c}{{Table 1}} \\
    \multicolumn{8}{c}{Summary of Derived Physical Characteristics of the Molecular Gas} \\
        \hline \hline
        Region & Species & Band wavelength & {Component} & $T_\mathrm{ex}$ & $N_\mathrm{tot}$ & $V_\mathrm{rad}$ & $\sigma_V$ \\
        & & ($\SI{}{\micron}$) & & ($\SI{}{\kelvin}$) & ($\SI{}{\centi\meter^{-2}}$) & ($\SI{}{\kilo\meter\per\second}$) & ($\SI{}{\kilo\meter\per\second}$) \\
        \hline  \hline
        AGN & CO & 4.7 & {Cold} & ${15 \pm 1}$ & ${\SI{{2.32 \pm 0.06}e17}{}}$ & ${-64.4 \pm 4.6}$ & - \\
        AGN & CO & 4.7 & {Warm} & ${205 \pm 141}$ & ${\SI{{6.79 \pm 3.27}e16}{}}$ & ${-57.0 \pm 10.6}$ & {-} \\
        \hline
        SF-NE & CO & 4.7 & {Cold} & ${32 \pm 4}$ & ${\SI{{1.76 \pm 0.16}e17}{}}$ & ${-7.0 \pm 1.2}$ & ${22.8 \pm 3.0}$ \\
        SF-NE & CO & 4.7 & {High-excitation} & ${668 \pm 28}$ & ${\SI{{1.55 \pm 0.06}e18}{}}$ & ${-0.3 \pm 3.0}$ & ${41.0 \pm 5.1}$ \\
        \hline
        SF-SW & CO & 4.7 & {Cold} & ${18 \pm 3}$ & ${\SI{{3.03 \pm 0.44}e17}{}}$ & ${-46.3 \pm 4.0}$ & ${68.2 \pm 6.0}$ \\
        SF-SW & CO & 4.7 & {Warm} & ${233 \pm 14}$ & ${\SI{{1.80 \pm 0.09}e18}{}}$ & ${-69.5 \pm 3.2}$ & ${85.8 \pm 4.7}$ \\
        SF-SW & CO & 4.7 & {High-excitation} & ${761 \pm 58}$ & ${\SI{{9.02 \pm 1.66}e17}{}}$ & ${-145.8 \pm 3.5}$ & ${47.9 \pm 5.3}$ \\
        \hline
        SF-NE & \ce{H2O} & 6.2 & & $298 \pm 71$ & $\SI{{9.3 \pm 1.8}e17}{}$ & $-2.3 \pm 4.4$ & $56.7 \pm 6.5$ \\
        \hline
        SF-SW & \ce{H2O} & 6.2 & & $191 \pm 37$ & $\SI{{9.5 \pm 2.4}e17}{} $ & $-107.6 \pm 6.1$ & $93.0 \pm 12.6$ \\
        \hline
        SF-NE & \ce{C2H2} & 13.7 & & $224 \pm 37$ & $\SI{{4.17 \pm 0.48}e15}{}$ & - & - \\
        \hline
        SF-NE & HCN & 14.0 & & $151 \pm 33$ & $\SI{{7.59 \pm 0.87}e15}{}$ & - & - \\
        \hline
    \end{tabular}
    \caption{Summary of the derived physical characteristics of all considered species and regions. The excitation temperature $T_\mathrm{ex}$, total species column density $N_\mathrm{tot}$, mean radial velocity $V_\mathrm{rad}$ and mean (instrument-corrected) velocity dispersion $\sigma_V$ are listed separately for each temperature component. For \ce{C2H2} and HCN, the radial velocity and dispersion are not constrained by the Q-branch fits; for the CO cold components, some lines are unresolved. All values were derived assuming a complete coverage of the background continuum source (see text).}
    \label{tab:results_summary}
\end{table*}

\section{Discussion}
\label{sec:discussion}

\subsection{Excitation mechanism} \label{sec:discussion-excitation}

The rotation diagram for CO in the SF-NE core (Fig.~\ref{fig:co_rot_diagram}) shows a well-constrained straight line for $J \gtrsim 7$, implying that the gas is in equilibrium at a rotational temperature of $\SI{700}{\kelvin}$. 
Remarkably, the gas appears to be fully thermalised out to at least $J=23$. 
These extremely high levels of thermalisation put very strong requirements on local densities if the levels are collisionally excited.
We therefore need to carefully consider the excitation mechanism at play. 
In this discussion, we focus on CO first, because for this molecule we have characterised the level populations in the most detail.

\subsubsection{Collisional excitation} \label{sec:discussion-excitation-collisional}

Collisional LTE occurs if collisional deexcitation dominates over radiative downward transitions, i.e., if the density of the main collisional partner species---typically H$_2$ for neutral molecules---is much greater than the relevant critical density.



To establish whether the CO rotation diagrams in Fig.~\ref{fig:co_rot_diagram} can be produced by collisional excitation, we estimate the critical density required to thermalise the highest levels that follow the straight line in the rotation diagram. For the SF-NE core, the levels appear to be
thermalised up to at least $J = 23$ ($E/k \approx \SI{1500}{\kelvin}$)---and likely up to $J = 27$---at a best-fit rotational temperature of $T_\mathrm{rot} \approx \SI{700}{\kelvin}$.
At the SF-SW position, the highest-excitation CO component displays LTE conditions at a temperature of ${T_\mathrm{rot} \approx \SI{800}{\kelvin}}$ up to at least ${J = 30}$ (${E/k \approx \SI{2600}{\kelvin}}$).

For the rovibrational transitions, the critical densities are extremely high, as the Einstein $A$ coefficients are of the order $A_\text{CO,rovib} \sim \SI{10}{\per\second}$, while the collisional de-excitation rates are at most $k_\text{CO,rovib} \sim \SI{e-17}{\centi\meter^{-3}\per\second}$. Thus, if collisions dominate the excitation of CO, it must be through the pure rotational transitions. Therefore, we consider the critical densities of rotational transitions in the vibrational ground state.

For the $J = 24-23$ transition of CO, the Einstein $A$ coefficient is $A_{24,23} = \SI{1.281e-3}{\per\second}$, while the sum of the collisional de-excitation rates amounts to $\sim \SI{5e-10}{\centi\meter\cubed\per\second}$. 
Thus, in the absence of a radiation field, the critical density is $n_\mathrm{crit,min} (J=24-23; T = \SI{700}{\kelvin}) \sim \SI{3e6}{\centi\meter^{-3}}$. In order for the gas to be thermalised up to this level, as the rotation diagram suggests, we therefore need densities $n(\text{H}_2) \gtrsim \SI{e7}{\centi\meter^{-3}}$. We combined non-LTE modelling using DESPOTIC \citep{despotic} with MCMC fitting using \texttt{emcee} \citep{emcee} to test whether the level populations that are not dominated by foreground gas ($J \geq 3$) in Fig.~\ref{fig:co_rot_diagram} could be produced in the absence of a radiation field. {In these models, statistical equilibrium level populations were calculated, considering rotational transitions with Einstein A coefficients and collisional rates taken from LAMDA \citep{LAMDA, Yang2010} and using the escape probability formalism assuming a spherical geometry.} The resulting posteriors suggest a similar density requirement of ${n(\mathrm{H_2}) \gtrsim \SI{6e6}{\centi\meter^{-3}}}$.


Although this density is not impossible on small scales such as the central regions of actively star-forming cores, such regions are expected to be sub-pc in size. In the present case however, the thermalised high-temperature components dominate the column densities over $R \approx \SI{130}{\parsec}$ apertures. {The absorption signal arises from a likely smaller region set by the projected area of the background continuum source, but radio observations show structures on the scale of several pc.} 
{\citet{Song2022} estimate deconvolved radii of $\approx \SI{40}{\parsec}$ for SF-NE and SF-SW from their $\SI{33}{\giga\hertz}$ VLA observations. ALMA observations of the \SI{350}{\giga\hertz} continuum also reveal structures that are considerably larger than the 0\farcs2 (\SI{80}{\parsec}) beam \citep{Saito2018}.} 
\citet{Iono2013} used HCN(4-3)/(1-0) and HCO$^+$(4-3)/(1-0) line ratios to estimate densities of $n(\text{H}_2) = \SI{e5}{}-\SI{e6}{\centi\meter^{-3}}$, which is still insufficient to thermalise the CO out to $J = 23$.

Therefore, it is highly unlikely that the CO we observe towards the SF-NE and SF-SW cores is actually in collisional LTE,
{unless the $\SI{4.7}{\micron}$ continuum totally originates in small ultradense ($n(\ce{H2})>\SI{e7}{\per\centi\meter\cubed}$) regions. However, even then collisional excitation will be irrelevant, since at such high densities the gas temperature will equal the dust temperature, which is set by the local radiation field.}

{Mapping of the [FeII] $\SI{5.34}{\micron}$ emission line has revealed the presence of a large-scale bow shock covering the SF-SW region \citep{Donnan2023}. It is of interest to consider the possible role of this shock for the excitation of the molecular gas in SF-SW, since C-type shocks in molecular clouds can easily produce post-shock temperatures of several times $\SI{e3}{\kelvin}$ without dissociating the molecules \citep{Draine1983}. In the present case however, the highly-excited molecular gas observed in absorption cannot be identified with this post-shock gas, as the \ce{H2O} and high-$J$ CO lines are blueshifted by $> \SI{100}{\kilo\meter\per\second}$ with respect to the [FeII] line. Furthermore, the detection of prominent \ce{CO2} ice absorption features (Fig.\ \ref{fig:nirspec_spectra_full}) shows that no significant grain mantle destruction, which always occurs in such shocks, has taken place. In addition, shocks produce a range of temperatures in the cooling post-shock gas, while the observed CO lines are well-represented by a single high-temperature component for $19 \leq J \leq 31$ (Fig. \ref{fig:co_rot_diagram}). We therefore conclude that the large-scale bow shock traced by [FeII] is not responsible for the high CO excitation in SF-SW, and that the role of more localised shocks is likely minor at most.}

{\citet{Donnan2023} also found strong [FeII] emission towards the SF-NE region. Here the molecular lines are all at rest with the [FeII] line, so an association between the two cannot be ruled out. However, strong absorption by \ce{CO2} ice is clearly present, and in this region the CO gas appears to be completely dominated by a single excitation temperature for $7 \leq J \leq 27$ (Fig. \ref{fig:co_rot_diagram}). For these reasons, shock excitation is unlikely here.}


\subsubsection{Far-IR radiative excitation}
\label{sec:discussion-excitation-farIR}


We next consider radiative excitation by the far-IR radiation field. This model is suggested by the fact that the relevant CO pure rotational lines have frequencies between 
$\SI{1037}{\giga\hertz}$ (\SI{289}{\micron}) for $J=9-8$---where the $\SI{700}{\kelvin}$ component begins to dominate---and $\SI{3098}{\giga\hertz}$ (\SI{97}{\micron}) for $J=27-26$. At these far-IR wavelengths, U/LIRGs have intense local radiation fields, and can become optically thick (resulting in a blackbody local radiation field) down to wavelengths of typically $\sim\SI{100}{\micron}$ \citep{Solomon1997}.
Since radiative excitation of the CO rotational lines has hardly been discussed in the literature, we summarise the relevant equations in Appendix~\ref{appendix:two-level-system}.

To equilibrate all the rotational levels probed, the radiation field must have the shape of a blackbody of at least the excitation temperature of the gas, across the indicated wavelength range.
A blackbody of $T = \SI{700}{\kelvin}$ peaks at $\SI{4.1}{\micron}$; one of $T = \SI{800}{\kelvin}$ peaks at $\SI{3.6}{\micron}$. However, the observed spectra of SF-NE and SF-SW peak at much longer wavelengths, as inspection of Fig.~\ref{fig:mrs_spectra_full} immediately confirms. {Furthermore, at the column densities \citep{Saito2015} and local extinctions \citep{Donnan2023} estimated for SF-NE and SF-SW, the dust will be optically thin at the wavelengths of the relevant CO lines.}
We conclude that radiative excitation by the far-IR radiation field must be ruled out as an excitation mechanism.

\subsubsection{Mid-IR pumping} \label{subsec:mid-IR-excitation}

Finally, we consider excitation by mid-IR radiation, through the vibrational transitions at approximately $\SI{4.7}{\micron}$. This mechanism is attractive since, as noted in Section~\ref{sec:discussion-excitation-farIR}, the $\sim\SI{700}{\kelvin}$ radiation field required to produce the observed level of radiative excitation, if interpreted as a blackbody, peaks at $\sim\SI{4}{\micron}$, so a strong radiation field is available. Close inspection of the CO band spectra presented in Fig.~\ref{fig:CO_5micron_spec} and Fig.~\ref{fig:co_model_spec_sw} reveals the presence of weak vibrational emission lines, in addition to the much stronger absorption lines, at several positions. As noted in Section~\ref{sec:discussion-excitation-collisional}, critical densities for the vibrational transitions are extremely high, so these emission lines cannot result from collisional excitation. Their presence therefore directly shows that radiative excitation through the $\SI{4.7}{\micron}$ band plays a role.

An additional argument supporting the importance of mid-IR pumping is provided by the pronounced observed asymmetry between the P- and R-branches of both the CO and the H$_2$O lines, which was noted in Sections~\ref{subsec:analysis-co} and \ref{sec:analysis-h2o}. In a situation where only absorption occurs, this asymmetry is impossible to explain, since the ratio of P- and R-branch lines originating from the same lower level is determined solely by the relevant Einstein $A$ coefficients. In the presence of radiatively excited emission this situation changes.
Consider a $\SI{4.7}{\micron}$ continuum photon that is absorbed in the R$(J')$ line, exciting a CO molecule to the $(v, J) = (1, J'+1)$ state. This molecule can radiatively decay back to the $v = 0$ state through either the R$(J')$ line or the P$(J'+2)$ line, with approximately equal probabilities.
However, since for any positive excitation temperature
$n_{J'}/g_{J'} > n_{J'+2}/g_{J'+2}$, absorption in the R($J'$) line will be stronger than that in the P($J'+2$) line. This mechanism, which was analysed by \citet{Gonzalez-Alfonso2002}, is illustrated in Fig.~\ref{fig:branch_asymmetry_diagram}. The stronger absorption in the R-branch line compared to the P-branch, while the emission lines have similar strengths, leads to the observed asymmetry between P- and R-branches. This process also results in a net population transfer to higher rotational levels, and may therefore affect the rotational temperature in the vibrational ground state \citep[see also][]{CarrollGoldsmith1981, Sakamoto2010}.

\begin{figure}[h!]
    \centering
    \includegraphics[scale=0.39]{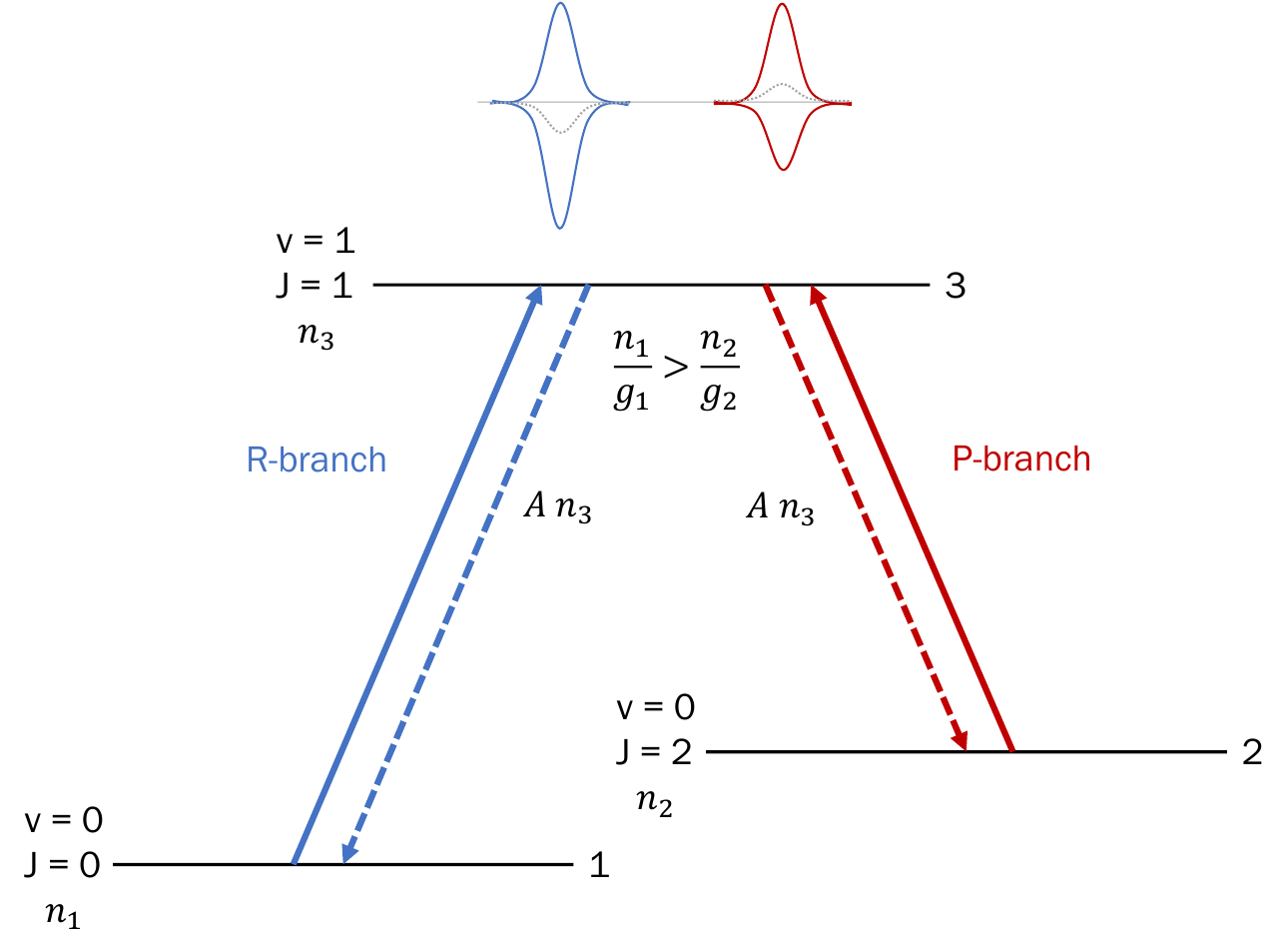}
    \caption{Schematic representation of the rovibrational excitation process in the case of a P- and R-branch. {The density of each level $i$ is denoted by $n_i$; $A$ denotes the Einstein A coefficient, which does not vary strongly between adjacent rovibrational lines.} {As long as there is no population inversion between levels 1 and 2 (i.e. $n_1/g_1 > n_2/g_2$), the R-branch absorption is stronger than the P-branch absorption, while the corresponding emission lines---both proportional to $A n_3$---are equally strong.} The net effect is 
    an asymmetry between the appearance of the P- and R-branches.}
    \label{fig:branch_asymmetry_diagram}
\end{figure}


Depending on the relative strength of emission and absorption, the asymmetry can lead to a situation where both branches are in absorption (as in the cases studied in this paper), or both are in emission, or where the R-branch is in absorption, but the P-branch appears in emission (as illustrated in Fig.~\ref{fig:branch_asymmetry_diagram}). An example of the latter situation is given by \citet[][their Fig.~3, middle panel]{Pereira-Santaella2023} {in the nearby LIRG NGC\,3256}. We have explored our datacube, and likewise find extended regions where the R-branch is seen in absorption and the P-branch in emission, confirming the importance of mid-IR radiative excitation.


If mid-infrared pumping indeed dominates the level populations in the vibrational ground state, the implications for the interpretation of the rotational temperature are profound. Since the rovibrational lines, unlike the pure rotational lines, all lie at effectively the same wavelength, a blackbody is not required to produce equilibrium level populations. Instead, the excitation temperature traces the almost monochromatic brightness temperature of the exciting radiation field at the wavelength of the transitions. Thus, the excitation temperature found for CO can be interpreted as a measure of the intensity of the local radiation field at $\SI{4.7}{\micron}$, and is in fact its (Planck) brightness temperature (see Appendix~\ref{appendix:two-level-system}). 

The above discussion assumes that the mid-IR radiation field not only excites the vibrational levels, but also determines the {high-$J$} rotational levels in the ground vibrational state. If the excitation is purely by radiation, this is guaranteed to be true, since in this case radiative equilibrium at the excitation temperature $T_\mathrm{rad}$ applies. However, collisions, if sufficiently frequent, can modify the populations of the rotational states. This situation was found in the extended outflow regions in NGC\,3256 studied by \citet{Pereira-Santaella2023}. In contrast, here we are studying the most intense luminous cores, where local radiation fields are much more intense. We evaluate the relative importance of radiative and collisional excitation of the rotational levels in Appendix~\ref{appendix:rotational_population}. This calculation shows that, in the intense mid-IR radiation fields encountered in the regions studied here, radiative pumping through the mid-IR vibrational levels also dominates the population of the rotational levels. Only for the lowest rotational levels, where collisional rates are significantly higher, collisional excitation may play a role, and this may account for the low excitation temperatures derived at levels well below $J=10$ (see Figs.~\ref{fig:co_rot_diagram} and \ref{fig:co_rot_diagram_sw} and Table~\ref{tab:results_summary}). {More complex modelling of the CO and H$_2$O bands by \citet{Gonzalez-Alfonso2023} supports this conclusion.}

Turning now to the other molecules, the same branch asymmetry that we noted for CO, was found for H$_2$O (see Section~\ref{sec:analysis-h2o}), confirming that radiative excitation plays a role. Furthermore, the quality of the LTE fit (see Fig.~\ref{fig:h2o-model-spec}) suggests equilibrium conditions, despite the high rotational critical densities of $n_\mathrm{crit} \sim \SI{e8}{\centi\meter^{-3}}$ for this molecule. Thus, the H$_2$O observed towards the SF-NE core is likely radiatively excited as well.

For HCN and C$_2$H$_2$ we can only analyse the Q-branch and the S/N of our data is not sufficient to conclude whether a branch asymmetry is present in these cases or not. Due to the limited S/N and spectral resolution, it is difficult to assess whether non-LTE effects play a role. We note, however, that HCN has high rotational critical densities ($n_\mathrm{crit} \sim \SI{e7}{}-\SI{e8}{\centi\meter^{-3}}$) and is strongly {susceptible to} both far-IR radiative excitation and mid-IR pumping \citep{Sakamoto2010}. Given the high \SI{14}{\micron} intensity observed towards the SF-NE core (Fig.~\ref{fig:mrs_spectra_full}), it is likely that HCN is excited through the mid-IR radiation field as well. For C$_2$H$_2$ no critical densities are available, and we cannot with certainty establish the dominant excitation mechanism.

{We note that the wavelengths of the molecular bands all lie close to the peak of a blackbody radiator at the excitation temperature derived for that band. This result is natural in the case of radiative excitation through a molecular band, since this mechanism is most energy-efficient if the exciting radiation field peaks close to the wavelength of that band. In the case of collisional excitation however, the lowest excitation temperatures would be expected for H$_2$O, which has the highest critical densities. The observed trend in excitation temperatures thus provides further support for radiative excitation through the vibrational bands.}

\subsection{AGN vs.\ starburst}   \label{sec:discussion_AGN_SB}

We have established that the observed molecular bands are dominated by radiative excitation. This conclusion provides an explanation for the conundrum described in Section~\ref{sec:discussion-excitation}: why do the molecular bands display much higher excitation in the star forming regions than at the AGN position?

The crucial observation is that the radiation field that provides the excitation is the \textit{angle-averaged} radiation field (see Eq.~\ref{eq:Draine_n_gamma} in Appendix~\ref{appendix:two-level-system}).
This leads to a pronounced difference in the excitation between the star forming regions and the AGN environment, which is illustrated in Fig.~\ref{fig:geometry-cartoon}. In the star forming regions, the exciting radiation field is provided by hot dust, heated by the young stars that form throughout the regions (plus possible a contribution from starlight). {This geometry is found for instance in the eponymous hot core in the Orion-KL region, which is externally heated by the surrounding young star cluster \citep{Blake1996,Zapata2011}.} Since the gas is mixed with the hot dust, there is no geometric dilution of the exciting radiation field. The core of the star-forming region will produce the most intense radiation, allowing gas in front of it---which is the gas probed by our data---to be observed in absorption. Hence, the \textit{local} mid-IR radiation field excites the molecular gas, and the more intense radiation from deeper layers of the core provides a background continuum.

The situation is different in the AGN case: the $\SI{4.7}{\micron}$ radiation field is now provided by the (parsec-size) circumnuclear torus, which serves both as a background source and provides the exciting radiation field. The solid angle from which the absorbing cloud is excited now depends on the size of the torus and the distance of the cloud from the torus, but is in any case much smaller than $4\pi\,$sr. As a result, the exciting radiation field 
is strongly geometrically diluted at the location of the absorbing gas. The result is a strongly reduced angle-averaged exciting radiation field, {and therefore the CO does not reach the high degree of excitation observed towards the intense star-forming regions.}

{As the CO line detections towards the AGN only extend to $J = 8$, for which the critical densities are only $n_\mathrm{crit} \sim \SI{e5}{\centi\meter^{-3}}$, and the rotational temperature is relatively low, we cannot definitively establish whether the $T_\mathrm{ex} \approx \SI{200}{\kelvin}$ component is dominated by collisions or radiation. We note, however, that the detection of rovibrational CO emission lines confirms that a substantial amount of CO near the AGN is infrared-pumped to the $v=1$ state, and therefore the pumping rate must be considerable.}

Intense local infrared radiation fields can also be produced very close to single massive, dust-embedded stars. CO located in the centrally heated dusty envelope will be exposed to a strong locally generated $\SI{4.7}{\micron}$ radiation field from every direction, if the region is optically thick at that wavelength. While this appears to be the case for SF-NE and SF-SW, we note that the AGN environment must have moderate $\SI{4.7}{\micron}$ optical depth, since the AGN is detected already at $\SI{2.3}{\micron}$, and is not associated with an extinction peak \citep[their Fig.~4]{Donnan2023}. Any CO in the AGN envelope will then be exposed to a much weaker $\SI{4.7}{\micron}$ radiation field than near the more deeply embedded stars, since the envelope is now optically thin. The physical principle of this model is the same as above, and only the adopted geometry is different.

{We finally note that the column density implied by the CO data towards the AGN is more than an order of magnitude lower than typical column densities of the geometrically thick AGN-obscuring circumnuclear torus \citep[$N(\ce{H2})\sim\SI{e23}{\per\centi\meter\squared}$, e.g.][]{HonigKishimoto2010}. Such a torus could only produce the low column densities inferred if it were seen nearly face-on, which would result in a type~1 AGN, contrary to what is observed. It is therefore not possible to associate this gas with the torus itself.}

\begin{figure}[h!]
    \centering
    \includegraphics[scale=0.35]{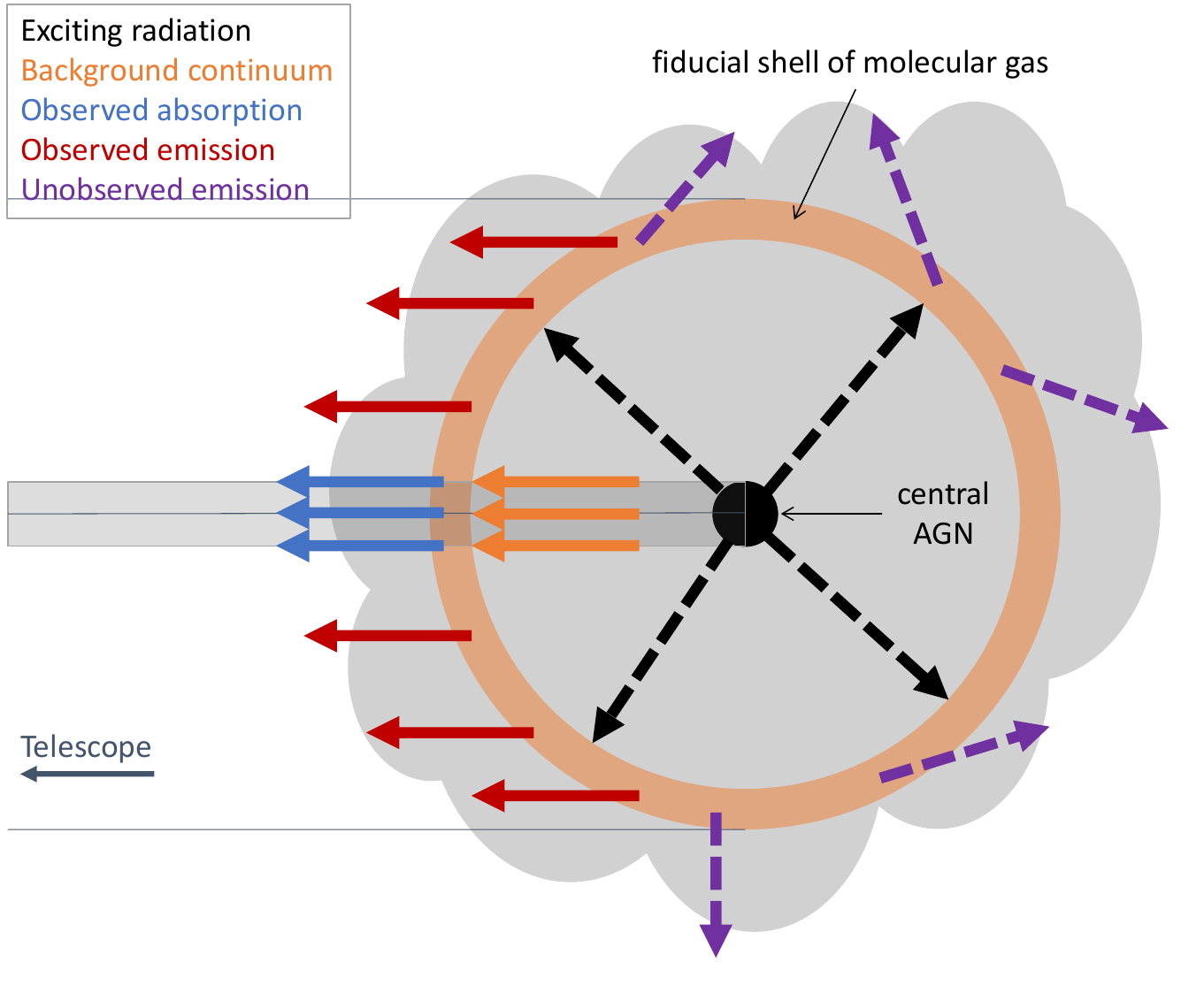}
    \includegraphics[scale=0.35]{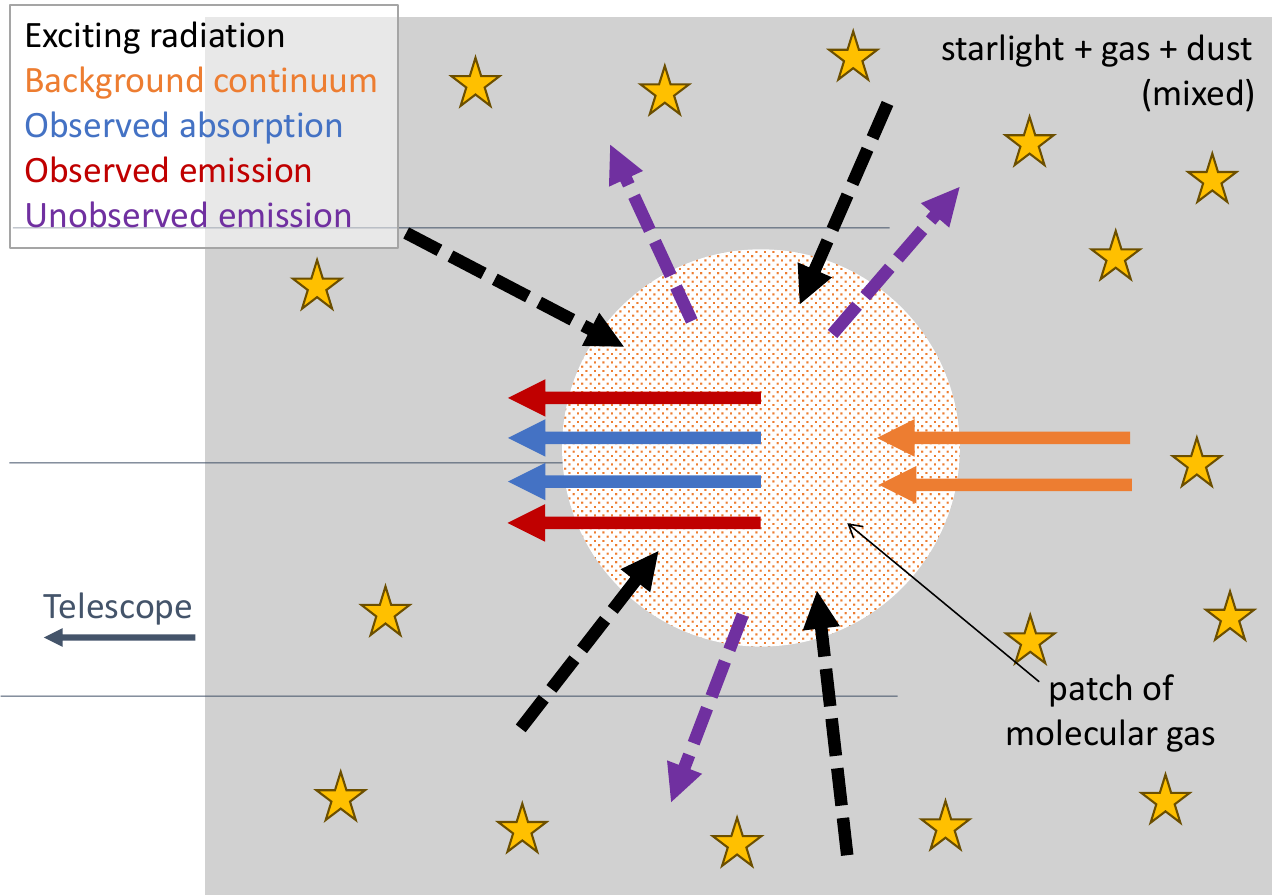}
    \caption{Schematic illustration of the geometry and excitation around an AGN (top) and in a star-forming knot (bottom). Solid arrows indicate observed radiation; dashed arrows illustrate undetected radiation. For the AGN region, molecular gas in a fiducial shell (orange) surrounding the central AGN + torus (black) is irradiated from only a small solid angle towards the AGN. Therefore, the molecular gas will not be highly-excited, despite the high intensity of the continuum emission from the AGN. In the case of a star-forming region, the gas, dust and starlight are effectively mixed, and therefore the sketched patch of molecular gas is irradiated isotropically by infrared continuum emission.}
    \label{fig:geometry-cartoon}
\end{figure}

\subsection{Local conditions in the star-forming regions}



{For SF-NE we have derived a maximum possible CO column density $N(\mathrm{CO})=\SI{6e18}{\per\centi\meter\squared}$. The local gas density as derived from CO at this position is about $\SI{e5}{\per\centi\meter\cubed}$, but considering also HCN and HCO$^+$ yields $\log n_{\mathrm{H}_2}=5.0-5.9$ \citep{Saito2015}. Combining these estimates gives a typical path length through the high-excitation CO component of at most $\SI{0.06}{pc}$. This small typical dimension is not surprising, since the required exciting radiation field can only be generated at the cores of the densest star clusters or close to very bright stars. The emerging picture thus has luminous dense clusters of young stars distributed throughout the star-forming regions SF-NE and SF-SW, with the high-excitation CO located in regions of high radiation intensity within such clusters, as illustrated in Fig.~\ref{fig:geometry-cartoon}.}

{In order to estimate whether the required luminosity densities can be achieved, we adopt parameters of well-studied super star clusters. The densest young massive star cluster in the Milky Way is the Arches cluster with a central density of approximately $\SI{2e5}{\Msun\per\parsec\cubed}$ \citep{Espinoza2009}. Rapid dynamical evolution will lead to even higher central densities \citep{Harfst2010}. A central density as high as $\SI{e7}{\Msun\per\parsec\cubed}$ has been proposed for the R136 cluster in the Large Magellanic Cloud, although this number is not undisputed \citep{SelmanMelnick2013}. In any case it seems not unreasonable to suppose that the massive young star clusters in SF-NE and SF-SW can reach central densities as high as about $\SI{e6}{\Msun\per\parsec\cubed}$. Using \texttt{Starburst99} \citep{Leitherer1999}, we find that an instantaneous starburst with a Salpeter Initial Mass Function (IMF) cut off at $\SI{100}{\Msun}$ has a bolometric $L/M$ ratio of about $\SI{1.7e3}{\Lsun\per\Msun}$ during the first few million years. Using simple energy density arguments, we can estimate the typical radiation intensity within these clusters and find that it falls short of the required value by about an order of magnitude.}

{However, this estimate is probably overly conservative, for several reasons. First, the star formation rates of SF-NE and SF-SW are much higher than the most extreme values found in Local Group galaxies; therefore, the clusters in SF-NE and SF-SW are likely significantly more extreme in mass, central density and stellar content than the densest young clusters in the Local Group, in particular if they undergo rapid dynamical evolution. Second, the IMF of the \texttt{Starburst99} models used above does not consider stars more massive than $\SI{100}{\Msun}$, while in reality the IMF in these regions is expected to be fully populated, extending well above $\SI{100}{\Msun}$. Finally, it has been proposed that in the most actively star-forming regions of (U)LIRGs the IMF may be top-heavy \citep{Romano2017,Sliwa2017,Zhang2018}. All of these effects would lead to a significantly more intense local radiation field than in the most extreme star clusters in Local Group galaxies.}
Indeed, ``Super Hot Cores'' associated with the youngest super star clusters have been found in the nearby starburst galaxy NGC\,253 \citep{RicoVillas2020}.

Nevertheless, it is clear that the required luminosity densities are extreme. We therefore also investigate the alternative model, where the molecular gas is simply located close to luminous young stars. Using the models for dust-embedded stars by \citet{ScovilleKwan1976}, we find that a $\SI{120}{\Msun}$ star, with a luminosity of $\SI{2e6}{\Lsun}$ \citep{Ekstrom2012}, can produce the required radiation intensity out to a distance of about $\SI{0.005}{pc}$. Assuming that the high-excitation CO is located in a layer of this thickness then implies a local density $n_{\mathrm{H}_2}\sim\SI{e6}{\per\centi\meter\cubed}$, in excellent agreement with observationally estimated values. Since the {most massive} stars form in dust-embedded ultracompact and hypercompact HII regions, where most of the Lyman continuum is absorbed by dust \citep[and references therein]{Churchwell2002,Hoare2007}, the molecular gas can be shielded from rapid destruction in these harsh environments.

{In summary, although the implied local radiation intensities are very high, they can still be produced by stellar sources in the regions of intense star formation, or close to the most luminous young stars, that are expected to exist in LIRGs, and our spectra provide ideal probes of local conditions in such regions. We note that an alternative model for generating the intense local radiation field, involving an accreting intermediate-mass black hole, has been suggested by \citet{Gonzalez-Alfonso2023}. In this case a compact non-thermal radio point source is expected to be present. Radio VLBI observations will thus be able to distinguish between these two models.}

\subsection{Kinematics and the role of shocks} \label{subsec:kinematics}


{The \ce{CO} spectrum observed towards the AGN (Fig.~\ref{fig:co_model_spec_sw}) shows P-Cygni profiles for the P(4)-P(8) lines: emission from infrared-pumped \ce{CO} gas is observed at the rest-frame wavelength, and blueshifted lines with a radial velocity of $V_\mathrm{rad} \approx \SI{-50}{\kilo\meter\per\second}$ with respect to \ce{H2} are seen in absorption. Absorption in these lines traces the ``warm" $T_\mathrm{ex} \approx \SI{200}{\kelvin}$ component (see Fig.~\ref{fig:co_rot_diagram_sw} and \ref{fig:co_model_spec_sw}) closer to the AGN. Thus, this outflow is directly associated with the nucleus, and these observational signatures can be produced by an expanding shell around the AGN. }

{Although the blueshifted absorption lines reveal the presence of outflowing moleculas gas, they only hold information {about the line-of-sight velocity and column density of the} the gas inside a narrow line of sight against the AGN background continuum. {An estimate of the mass outflow rate is precluded by the absence of information on the dimensions and geometry of the outflow.} 
The P-branch emission lines originate from a larger projected area around the AGN\null. However, due to the partial blending with the absorption lines, we cannot reliably fit their line profiles, and therefore we cannot robustly estimate the corresponding column densities or velocity dispersions. Thus, further characterisation of the \ce{CO} outflow is {not possible with the present data}.}


{While the excitation conditions of the observed molecules in the SF-NE and SF-SW star-forming regions are similar, we find significant differences in the kinematics. Towards the SF-NE region, the absorption lines of \ce{CO}, \ce{H2O}, \ce{C2H2} and \ce{HCN} are all effectively at rest with the bulk of the warm molecular gas, as traced by \ce{H2}. The SF-SW region, however, exhibits strong blueshifts and generally broader lines for both \ce{CO} and \ce{H2O}. The cold foreground \ce{CO} observed towards this region has a similar radial velocity as that observed towards the nearby AGN; the more highly excited \ce{CO} reaches a radial velocity of $V_\mathrm{rad} \approx \SI{-150}{\kilo\meter\per\second}$. The \ce{H2O} is strongly blueshifted as well, with $V_\mathrm{rad}\approx\SI{-100}{\kilo\meter\per\second}$.} 
{These findings indicate that, in the SF-SW core, the radiatively-excited gas close to the young stars is driven out by stellar winds or radiation pressure.} 


\subsection{Relation to Galactic star forming regions, ULIRGs and CONs}

{Since the highly-excited vibrational bands probe the environments of the most massive recently-formed stars, it is not surprising that the spectra show many of the features typically found in Galactic high-mass star forming regions, such as strong absorption features from gas-phase and solid-state species, and a rising continuum towards long wavelengths \citep[e.g.,][]{Evans1991,Boonman2003,BoonmanVanDishoeck2003,BarentineLacy2012,Li2022,Beuther2023}. However, closer inspection reveals a number of interesting differences. The most luminous dust-embedded young star known in the Milky Way is AFGL\,2591\,VLA\,3 with a luminosity of about $\SI{2.3e5}{\Lsun}$ and a mass of $\SI{40}{\Msun}$. Molecular vibrational bands between 4 and $\SI{13}{\micron}$ have been studied by \citet{Barr2020,Barr2022}, who find a fairly constant excitation temperature of $600-\SI{700}{K}$ for the warm gas component across a number of different molecular species. This result points to thermalised collisional excitation, possibly in a circumstellar disk, and indeed the estimated density $n_{\mathrm{H}_2}\sim\SI{e9}{\per\centi\meter\cubed}$ is much higher than the density estimates in VV\,114.}

{Another interesting comparison can be made with the externally-heated hot core in the Orion-KL region. It displays CO vibrational bands in absorption towards the object IRc2 \citep{Gonzalez-Alfonso2002}, which display a striking similarity to the bands observed in SF-NE and SF-SW\null. These and more recent observations \citep{Beuther2010,Nickerson2023} show an excitation temperature of $100-\SI{200}{K}$ and a density of a few times $\SI{e8}{\per\centi\meter\cubed}$. It thus appears that conditions in SF-NE and SF-SW correspond to the highest excitation temperatures found in Galactic high-mass star forming regions but do not require the very high local densities often invoked for these regions.}

{The average infrared surface luminosity of SF-NE is about $\SI{e13}{\Lsun\per\kilo\parsec\squared}$. This number is typical also of ULIRGs \citep{Thompson2005}. Therefore we may expect in ULIRGs local conditions similar to those derived in the present paper. Our results thus confirm earlier suggestions that the {star formation rate surface densities} 
of U/LIRGs are similar to those of the most active star formation regions in the Milky Way, but scaled up to the entire interstellar medium of these galaxies \citep{Solomon1997,DownesSolomon1998,Papadopoulos2012}. U/LIRGs display a large range of extinctions \citep[e.g.,][]{Spoon2007,Spoon2022}, and we may expect to be able to utilise the same excitation diagnostics as in the present paper in the low-extinction ULIRGs, but they may not be accessible in the more obscured systems.}

{Compact Obscured Nuclei (CONs) are defined by their luminous rotational HCN emission lines in the $\nu_2{=}1$ vibrational state \citep{Falstad2021}. They contain highly obscured ($N(\mathrm{H}_2)>\SI{e24}{\per\centi\meter\squared}$), compact ($10-\SI{100}{\parsec}$) and luminous nuclei and are found in about 50\% of all ULIRGs and 20\% of LIRGs. No $\nu_2=1$ HCN emission lines have been detected from VV\,114 \citep{Saito2018}, and the column densities derived for VV\,114 are much lower than typical CON values. Furthermore, the AGN identified in VV\,114 is only moderately obscured. However, the HCN absorption band discussed in the present paper does provide the population of the $\nu_2=1$ level. Studies of this band will thus provide a new diagnostic of local conditions in CONs, as long as foreground extinction is not so high that it prevents detection of the region where the $\nu_2=1$ lines originate.}

\section{Conclusions}
In this work, we have made use of JWST NIRSpec and MIRI MRS spectra in $R \approx \SI{130}{\parsec}$ apertures to study the rovibrational absorption spectra of CO, H$_2$O, C$_2$H$_2$ and HCN in the heavily-obscured nucleus of VV 114E. Two of the regions studied (SF-NE and SF-SW) have been previously identified as regions of intense star formation, and one region (labeled ``AGN" in this work, or ``SW core" in previous studies) has been reported as an AGN. We reach the following conclusions:

\begin{enumerate}
    \item The detection of CO bandhead absorption at \SI{2.3}{\micron} towards regions SF-NE and SF-SW, but not towards the AGN, is further evidence of the stellar origin of the former and AGN nature of the latter (Section~\ref{subsec:co_overtones}).
    \item We detect highly-excited CO in the star-forming regions, with excitation temperatures as high as $T_\mathrm{ex} \approx 700$ and $\SI{800}{\kelvin}$. Towards the AGN, we {do not detect such a high-excitation component, and} find only weakly-excited CO with $T_\mathrm{ex} \approx \SI{200}{\kelvin}$ (Section~\ref{subsec:analysis-co}).
    \item Strong absorption lines of gas-phase \ce{H2O} are detected towards the SF-NE and SF-SW regions, at excitation temperatures of $T_\mathrm{ex} \approx \SI{300}{\kelvin}$ and $\SI{200}{\kelvin}$ respectively. Towards the AGN, \ce{H2O} lines are detected, but much weaker (Section~\ref{sec:analysis-h2o}).
    \item We confirm the detection of \ce{C2H2} and \ce{HCN} towards the SF-NE region, and find excitation temperatures of $T_\mathrm{ex} \approx \SI{220}{\kelvin}$ and $\SI{150}{\kelvin}$ respectively. These absorption features are potentially seen towards the SF-SW region as well, but at much lower signal-to-noise (Section~\ref{subsec:analysis-14micron}).
    \item We observe 
    {an asymmetry between the P- and R-branches of the vibrational bands of both CO and \ce{H2O}, which is characteristic of radiative excitation of the vibrational levels.}
    Towards the AGN, we observe P-Cygni profiles in the CO P(4)-P(8) lines, indicative of a molecular outflow at radial velocity $V_\mathrm{rad} \approx \SI{50}{\kilo\meter\per\second}$ with respect to \ce{H2} lines (Sections~\ref{sec:analysis-h2o}, \ref{sec:analysis-h2o} and \ref{subsec:kinematics}).
    \item {We conclude that mid-infrared radiative pumping can best account for the equilibrium conditions at high rotational temperature found for CO in the star forming regions} (Section~\ref{sec:discussion-excitation}).
    \item 
    {We conclude that both the P-/R-branch asymmetry and the highly-excited CO are indicative of the geometry of a star-forming region, where molecular gas is mixed with the dust that radiatively pumps it to higher vibrational states.
    For gas surrounding the AGN, the radiation field is geometrically diluted, {leading to lower excitation temperatures} 
    (Section~\ref{sec:discussion_AGN_SB}).}
    \item {We find that in the star-forming regions, the CO vibrational bands probe the highest local radiation fields, associated with young super star clusters or the most massive young stars.}
\end{enumerate}

In this work we have demonstrated the power of JWST NIRSpec/MIRI MRS IFU observations of rovibrational bands to characterise the excitation conditions of molecular gas in highly-obscured U/LIRG nuclei, as well as the strong effects mid-infrared pumping may have on the molecular gas in intense starburst regions. Future studies of these molecular bands may explore this new avenue to potentially distinguish between pure starburst effects and AGN activity in similarly-obscured systems.

\begin{acknowledgments}
We thank Ewine van Dishoeck for useful discussions.

This work is based on observations made with the NASA/ ESA/CSA James Webb Space Telescope. The research was supported in part by NASA grant JWST-ERS-01328. The data were obtained from the Mikulski Archive for Space Telescopes at the Space Telescope Science Institute, which is operated by the Association of Universities for Research in Astronomy, Inc., under NASA contract NAS 5-03127 for JWST. The observations analysed here can be accessed via \dataset[10.17909/yqk1-jr92]{\doi{10.17909/yqk1-jr92}}.

S.V. acknowledges support from the European Research Council (ERC) Advanced Grant MOPPEX 833460.
H.I. acknowledges support from JSPS KAKENHI Grant Number JP21H01129.
The Flatiron Institute is supported by the Simons Foundation.
A.M.M. acknowledges support from the National Science Foundation under AAG \#2009416 and CAREER \#2239807 and the NASA Astrophysics Data Analysis Program (ADAP) grant \#80NSSC23K0750.
C.R. acknowledges support from the Fondecyt Regular grant 1230345 and ANID BASAL project FB210003.
{V.U acknowledges funding support from \#JWST-GO-01717.001-A, \#HST-AR-17065.005-A, \#HST-GO-17285.001-A, and NASA ADAP grant \#80NSSC20K0450.}
\end{acknowledgments}

\facility{JWST (NIRSpec and MIRI)}
\software{Astropy \citep{astropy:2013, astropy:2018, astropy:2022},
    APLpy \citep{aplpy},
    NumPy \citep{numpy},
    SciPy \citep{scipy}
    }

\bibliography{biblio}{}
\bibliographystyle{aasjournal}

\appendix
\section{Excitation and radiative transfer}

In this appendix, we discuss the excitation and radiative transfer concepts needed to interpret our results. We refer to \textit{Physics of the Interstellar Medium} by \cite{Draine_book} as D11.

\subsection{Rotation diagrams from absorption spectra}   \label{appendix:rot_diagrams}

We consider a homogeneous medium, observed towards a background continuum source with specific intensity $I_{\nu,\mathrm{cont}}$. The equation of transfer for this situation is:

\begin{equation} \label{eq:eqtra}
    I_\nu = I_{\nu,\mathrm{cont}}\,e^{-\tau}+B_\nu\left(T_\mathrm{ex}\right)\left(1-e^{-\tau}\right).
\end{equation}

Here $T_\mathrm{ex}$ is the excitation temperature of the transition considered, and $B_\nu$ denotes the Planck function.

To construct a rotation diagram from an absorption spectrum, it is assumed that there is no significant line emission, i.e., the second term on the right-hand side of Eq.~\ref{eq:eqtra} can be ignored. This requires that $I_{\nu,\mathrm{cont}}\gg B_\nu\left(T_\mathrm{ex}\right)$. As pointed out by \citet{Lacy2013}, this condition is not always satisfied, so it is necessary to verify its validity. In case the condition is not satisfied, only the net absorption is measured, and ignoring the emission term will lead to underestimated column densities. {In the present case we find maximum excitation temperatures of approximately $\SI{700}{\kelvin}$
. Since all R-branch CO lines are observed in absorption, the condition $I_{\nu,\mathrm{cont}}>B_\nu(\SI{700}{\kelvin})$ must be fulfilled at $\SI{4.7}{\micron}$. The actual value of $I_{\nu,\mathrm{cont}}$ at this wavelength cannot be determined from our measurements, since the emission sources are not spatially resolved. Instead we explore plausible ranges and estimate their effects on our results. The background continuum cannot exceed the level of a $1500{-}\SI{1800}{\kelvin}$ blackbody, since the dust sublimation temperature is in this range \citep{Barvainis1987,HonigKishimoto2010}. With such a background continuum, our column densities would be underestimated by $5{-}8\%$. For an underestimate by more than a factor of 2, the background continuum would have to correspond to a blackbody in the narrow temperature range of $700{-}\SI{830}{\kelvin}$.}

Keeping these caveats in mind, the observed intensity spectrum can directly be converted to an optical depth spectrum:

\begin{equation}
    \tau_\nu = -\ln \left(\frac{I_\nu}{I_{\nu, \mathrm{cont}}} \right)
\end{equation}
The optical depth of an absorption line is given by D11 (his Eq.~8.6):

\begin{equation} \label{eq:tau}
    \tau_\nu = \frac{A_{ul}}{8\pi} \lambda_{ul}^2 \varphi_\nu \frac{g_u}{g_l} N_l \left( 1 - e^{-hc/k \lambda_{ul} T_\mathrm{vib}} \right)
\end{equation}

Here $u$ and $l$ denote the upper and lower levels involved in the transition. $A_{ul}$ is the corresponding Einstein A coefficient, $g$ is the level degeneracy, and $N_l$ is the column density in the lower level. $\lambda$ is the wavelength and $\nu$ is the corresponding frequency. The line profile is denoted by $\varphi_\nu$, which is subject to the normalization condition $\int\varphi_\nu\,d\nu=1$. $T_\mathrm{vib}$ is the excitation temperature between levels $u$ and $l$; in rovibrational spectroscopy, it is a vibrational temperature as the upper level is in the vibrationally excited state.

The stimulated emission term (the 2nd term in the brackets in Eq.~\ref{eq:tau}) can usually be ignored; {we verify the validity of this approximation in the context of this work in Appendix~\ref{appendix:rotational_population}}. We can then directly relate the column density in the lower level and the optical depth of the line:

\begin{equation}
    \tau_\nu = \frac{A_{ul}}{8\pi} \lambda_{ul}^2 \varphi_\nu \frac{g_u}{g_l} N_l
\end{equation}

The column density $N_l$ can be determined from the integrated optical depth. We perform the integral in wavelength-space, as the mid-infrared spectra are typically presented on a wavelength axis.

\begin{align}
    N_l &= \frac{8\pi \nu_{ul}^2}{A_{ul}c^2} \frac{g_l}{g_u} \int \tau_\nu \mathrm{d}\nu \\
    &= \frac{8\pi c}{A_{ul} \lambda_{ul}^4} \frac{g_l}{g_u} \int \tau_\lambda \mathrm{d} \lambda
\end{align}

In practice, we fit Gaussian profiles to the wavelength-space lines, and use the analytical solution of the integral to determine the column densities. For an amplitude $\tau_0$ and a line width $\sigma_{\tau, \lambda}$, we compute the column density as:

\begin{equation} \label{eq:coldens_practical}
    N_l = \frac{8 \sqrt{2} \pi^{3/2} c}{A_{ul} \lambda_{ul}^4} \frac{g_l}{g_u} \tau_0 \sigma_{\tau,\lambda}
\end{equation}

If stimulated emission is actually important, the derived column densities $N_l$ can be (approximately) corrected by dividing them by a factor $1 - e^{hc/k\lambda_ul T_\mathrm{vib}}$.
\\

The derived column densities can now be used to construct a rotation diagram, where we plot the log of weighted column densities $\ln(N_l / g_l)$ as a function of the corresponding level temperature $E_l/k$.
If the gas is in local thermodynamic equilibrium (LTE) at an excitation temperature $T_\mathrm{ex}$, the levels will follow a Boltzmann distribution:

\begin{equation} \label{eq:rot_diagram_LTE}
    \ln \left( \frac{N_l}{g_l} \right) = \ln {N_\mathrm{tot}} - \ln {Z(T_\mathrm{ex})} - \frac{E_l}{k T_\mathrm{ex}}
\end{equation}

Here $N_\mathrm{tot}$ is the total column density of the species (as far as it is at this excitation temperature), and $Z(T)$ is the partition function. In the rotation diagram, a medium in LTE at excitation temperature $T_\mathrm{ex}$ will display a straight line with absolute slope proportional to $1/T_\mathrm{ex}$. If there are several distinct excitation temperature components along the line of sight, or if the medium is not in LTE, the measurements will deviate from this simple linear relation.
However, one may still infer one or several excitation temperatures to characterise the system.
\\

The rotation diagrams in this paper are constructed from rovibrational lines from levels in the vibrational ground state. Thus, the excitation temperatures found in these rotation diagrams are \textit{rotational temperatures}, $T_\mathrm{rot}$. The \textit{vibrational temperature} $T_\mathrm{vib}$ describes the distribution of vibrational levels, which cannot be directly determined from the fundamental band. For absorption data, the presence of hot bands would provide constraints on the vibrational temperature.


\subsection{Combined collisional and radiative excitation of a two-level system}   \label{appendix:two-level-system}

In order to best illustrate the physical concepts, we focus our description on a two-level system. The concepts and methods are easily generalised to multi-level systems. Our treatment follows largely the methods of D11.

The system can undergo collisional transitions, spontaneous radiative decay, and induced radiative transitions. We parameterise the radiation field through the \textit{angle-averaged} photon occupation number $n_\gamma$ and the radiation temperature $T_\mathrm{rad}$:

\begin{equation} \label{eq:Draine_n_gamma}
    n_\gamma = \frac{c^3}{8\pi h \nu^3} u_\nu = \frac{1}{e^{h\nu / kT_\mathrm{rad}} - 1}
\end{equation}

Here $u_\nu$ is the specific energy density of the radiation field.
Since $u_\nu = (4\pi/c)\,\bar{I}_\nu$, where $\bar{I}_\nu$ is the angle-averaged specific intensity, $n_\gamma$ simply expresses $\bar{I}_\nu$ in a dimensionless form. The radiation temperature $T_\mathrm{rad}$, which is defined by Eq.~\ref{eq:Draine_n_gamma}, represents the temperature that a Planck function would have, in order to generate a radiation field with energy density $u_\nu$. It thus expresses the radiation field as a (Planck) brightness temperature.
\\

Following D11 (his Eq.~(17.5)), the upper level (1) is now populated and depopulated as:

\begin{equation}
    \frac{dn_1}{dt} = n_0 \left[ k_{01} n_c + \frac{g_1}{g_0} A_{10} n_\gamma \right] - n_1 \left[ k_{10} n_c + A_{10} (1 + n_\gamma) \right]
\end{equation}

Here $k_{01}$ and $k_{10}$ are the collisional excitation and de-excitation coefficients, $A_{10}$ is the Einstein $A$ coefficient of the transition, $n_i$ and $g_i$ are the density and degeneracy of level $i$, and $n_c$ is the density of the dominant collision partner---H$_2$ in the case of molecular gas.
\\

In statistical equilibrium, the level populations are now:

\begin{equation} \label{eq:Draine_2level_populations}
    \frac{n_1}{n_0} = \frac{k_{01} n_c + \frac{g_1}{g_0} A_{10} n_\gamma}{k_{10} n_c + A_{10} (1 + n_\gamma)}
\end{equation}
\\

The critical density in the presence of radiation is expressed as

\begin{equation}
    n_\mathrm{crit} = \frac{(1 + n_\gamma) A_{10}}{k_{10}}
\end{equation}

In the absence of a radiation field, this expression reduces to the usual ratio of the Einstein $A$ coefficient and the collisional de-excitation coefficient. If there is a strong radiation field, the critical densities are increased due to the addition of stimulated emission; a higher density is needed to compete with the increased rate of radiative decay.
\\

The collisional excitation and de-excitation coefficients are related by

\begin{equation}
    k_{01} = k_{10} \frac{g_1}{g_0} e^{-E_{10}/kT_\mathrm{kin}}
\end{equation}

Here $E_{10} = E_1 - E_0$ is the energy difference between the two levels; $T_\mathrm{kin}$ is the kinetic temperature of the gas. We can use this relation to simplify equation \ref{eq:Draine_2level_populations} to:

\begin{equation}
    \frac{n_1}{n_0} = \frac{n_c}{n_c + n_\mathrm{crit}} \frac{g_1}{g_0} e^{-E_{10}/kT_\mathrm{kin}} + \frac{n_\mathrm{crit}}{n_c + n_\mathrm{crit}} \frac{g_1}{g_0} \frac{n_\gamma}{1 + n_\gamma}
\end{equation}

But from equation \ref{eq:Draine_n_gamma}, it follows that $n_\gamma / (1 + n_\gamma) = e^{-h\nu / kT_\mathrm{rad}}$. We can therefore express the level populations as:

\begin{equation} \label{eq:2level_pop_ratio_temperatures}
    \frac{n_1}{n_0} = \frac{1}{1 + \frac{n_\mathrm{crit}}{n_c}} \frac{g_1}{g_0} e^{-E_{10}/kT_\mathrm{kin}} + \frac{1}{1 + \frac{n_c}{n_\mathrm{crit}}} \frac{g_1}{g_0} e^{-E_{10}/kT_\mathrm{rad}},
\end{equation}
where we have used the fact that the radiation field involved is that at the transition energy $h\nu = E_{10}$. Equation~\ref{eq:2level_pop_ratio_temperatures} provides a completely general expression for the excitation of a 2-level system excited by both collisions and radiation. Which mechanism dominates the excitation is determined in the first place by the local density $n_c$: if $n_c\gg n_\mathrm{crit}$ and no strong radiation field is present, collisions dominate and the excitation temperature will approach $T_\mathrm{kin}$.
If on the other hand $n_c\ll n_\mathrm{crit}$, and a strong radiation field is present, radiative excitation will dominate and the excitation temperature will approach $T_{rad}$.
\\

If the populations are governed by the radiation field we obtain
\begin{equation} \label{eq:2level_pop_ratio_radiation}
    \frac{n_1}{n_0} \approx \frac{g_1}{g_0} e^{-h\nu/kT_\mathrm{rad}}.
\end{equation}
In other words, if the radiation field dominates the level populations, we retrieve a Boltzmann distribution at the radiation temperature, i.e., \textit{radiative LTE}\null. The excitation temperature of the gas is then equal to $T_\mathrm{rad}$.
If a radiation-dominated system has several lines at various widely spaced frequencies, a single excitation temperature will be found only if the exciting radiation field is a blackbody over that entire frequency range.


\subsection{Population of the rotational levels in the vibrational ground state}
\label{appendix:rotational_population}

Here we estimate the relative importance of radiative and collisional excitation of the rotational levels of the ground vibrational state. Denoting the particle density in level $j$ by $n_j$, the rate of collisional population $R_{j,\mathrm{coll}}$ (in units cm$^{-3}\,$s$^{-1}$) is given by
\begin{equation}\label{eq:collisional_rate}
R_{j,\mathrm{coll}}=n_{\mathrm{H}_2}\,\Sigma_i k_{ij}n_i,
\end{equation}
where $k_{ij}$ is the collisional excitation coefficient for $E_i<E_j$,
the collisional de-excitation coefficient for $E_i>E_j$, and $k_{ii}=0$.

The radiative population rate for level $j$ can be calculated as
\begin{equation}\label{eq:radiative_rate_0}
    R_{j,\mathrm{rad}}=A_\mathrm{rovib}\,(1+n_\gamma)\,\frac{n_{v=1,j+1}+n_{v=1,j-1}}{2},
\end{equation}
which takes into account both P- and R-branch transitions; the $n_\gamma$ term corrects for stimulated emission. This expression makes use of the fact that the Einstein coefficients $A_\mathrm{rovib}$ for rovibrational transitions depend only weakly on the rotational quantum number.
Since the population of the vibrational levels is purely radiative (see Section~\ref{sec:discussion-excitation}), we can write
\begin{equation}
    \frac{n_{v=1,j}}{n_j}=e^{-(E_{v=1,j}-E_j)/kT_{\mathrm{rad}}},
\end{equation}
and since the vibrationally excited levels in Eq.~\ref{eq:radiative_rate_0} are close enough in energy that the difference can be ignored, we obtain to sufficient accuracy
\begin{align}
    R_{j,\mathrm{rad}} &= A_\mathrm{rovib}\,(1+n_\gamma)\,\frac{n_{j+1}+n_{j-1}}{2}\,e^{-hc/\lambda kT_\mathrm{rad}} \\
    &= A_\mathrm{rovib} \,n_\gamma\, \frac{n_{j+1} + n_{j-1}}{2},\label{eq:radiative_rate}
\end{align}
where $\lambda$ is the central wavelength of the vibrational band.

In order to establish whether collisional or radiative excitation dominates the rotational levels, Eqs.~\ref{eq:collisional_rate} and \ref{eq:radiative_rate} can be evaluated using population ratios from Figs.~\ref{fig:co_rot_diagram} and \ref{fig:co_rot_diagram_sw}. We illustrate the process by considering
the $J=10$ level, which is one of the lowest levels that appears to be definitely in the high-excitation components. CO data from ALMA analysed by \citet{Saito2015}, imply $T_\mathrm{kin}=25-\SI{90}{\kelvin}$ and $\log\,n_{\mathrm{H}_2} = 3.5-5.0$ (valid for region R21 in their Table~12, which corresponds most closely to the {SF-NE region} studied here). We choose $T_\mathrm{kin}=\SI{90}{\kelvin}$, in order to calculate the maximum possible collisional population rate. With these parameters, collision coefficients from \citet{Yang2010}, and expressing all level populations relative to $n_9$ (for reasons that will become clear shortly), we find
\begin{equation}
R_{{10},\mathrm{coll}}\approx \SI{5e-5}{\per\second} n_9\,\frac{n_{\mathrm{H}_2}}{\SI{e5}{\per\centi\meter\cubed}}. 
\end{equation}

In order to evaluate Eq.~\ref{eq:radiative_rate}, we use the fact that $A_\mathrm{rovib}\sim\SI{10}{\per\second}$ for CO (from the HITRAN database; see \ref{subsec:analysis-co} for further references). With $T_\mathrm{rad}=\SI{700}{\kelvin}$ and $\lambda = \SI{4.67}{\micron}$, we find from Eq.~\ref{eq:Draine_n_gamma} that $n_\gamma=0.012$, so that stimulated emission can be ignored. We then obtain
\begin{equation}
    R_{{10},\mathrm{rad}}\approx \SI{0.06}{\per\second}\,(n_9+n_{11}).
\end{equation}
In this expression, $n_{11}$ can easily be expressed in terms of $n_9$ using the population ratios displayed in Figures~\ref{fig:co_rot_diagram} and \ref{fig:co_rot_diagram_sw}, which then allows the radiative and collisional population rates to be compared directly.

This calculation shows that in the strong radiation fields in the star-forming regions considered here, radiative excitation of the rotational levels dominates over collisional excitation of the high-temperature components by several orders of magnitude, even at the highest densities proposed for these regions.
\\




\end{document}